\documentclass[aps,prd,superscriptaddress,showpacs,amsmath,amssymb,amsfonts,nofootinbib,showpacs,reprint]{revtex4-1}

\pdfoutput=1

\usepackage{graphicx}
\usepackage{color}
\usepackage{slashed}
\usepackage{tabularx}
\usepackage{hyperref}
\usepackage{tkz-graph}

\newcommand{\tr}{\mathrm{tr}}

\newcommand{\Nf}{N_{\text{f}}}

\newcommand{\sgn}{\operatorname{sgn}}

\newcolumntype{L}[1]{>{\raggedright\arraybackslash}p{#1}} %linksbündig mit Breitenangabe
\newcolumntype{C}[1]{>{\centering\arraybackslash}p{#1}} %zentriert mit Breitenangabe
\newcolumntype{R}[1]{>{\raggedleft\arraybackslash}p{#1}} %rechtsbündig mit Breitenangabe

\graphicspath{{.}{figures/}{mma/}{figures}}

\definecolor{blue}{rgb}{0,0,1}

\definecolor{green}{rgb}{0,1,0}

\definecolor{red}{rgb}{1,0,0}

\begin{document}

\title{Continuum Goldstone spectrum of two-color QCD at finite density\\ with staggered quarks}

\newcommand{\JLU}{Institut f\"ur Theoretische Physik, Justus-Liebig-Universit\"at, Heinrich-Buff-Ring 16, 35392 Giessen, Germany}
\newcommand{\JGU}{Institut f\"ur Kernphysik, Johannes-Gutenberg-Universit\"at, Johann-Joachim-Becher-Weg 45, 55099, Mainz, Germany}

\author{Jonas Wilhelm}\affiliation{\JLU}\affiliation{\JGU}
\author{Lukas Holicki}\affiliation{\JLU}
\author{Dominik Smith}\affiliation{\JLU}
\author{Bj\"orn Wellegehausen}\affiliation{\JLU}
\author{Lorenz von Smekal}\affiliation{\JLU}

\begin{abstract}
We carry out lattice simulations of two-color QCD and spectroscopy at finite density with two flavors of rooted-staggered quarks and a diquark source term. 
As in a previous four-flavor study \cite{Kogut:2003ju}, for small values of the inverse gauge coupling we observe a Goldstone spectrum which 
reflects the symmetry-breaking pattern of a Gaussian symplectic chiral random-matrix ensemble (GSE) with Dyson index $\beta_D=4$, 
which corresponds to any-color QCD with adjoint quarks in the continuum instead of QC$_2$D wih fundamental quarks. 
We show that this unphysical behavior occurs only inside of the bulk phase of 
$SU(2)$ gauge theory, where the density of $Z_2$ monopoles is high. 
Using an improved gauge action and a somewhat larger inverse coupling to suppress these monopoles, 
we demonstrate that the continuum Goldstone spectrum of two-color QCD, corresponding to a Gaussian orthogonal ensemble (GOE) with Dyson index $\beta_D=1$, 
is recovered also with rooted-staggered quarks once simulations are performed away from the bulk phase. We further demonstrate how this change of  
random-matrix ensemble is reflected in the distribution of eigenvalues of the Dirac operator. By computing the unfolded level spacings inside and outside 
of the bulk phase, we demonstrate that, starting with the low-lying eigenmodes which determine the infrared physics, the distribution of eigenmodes 
continuously changes from the GSE to the GOE one as monopoles are suppressed. 
\end{abstract}

%\keywords{two-color QCD}

%12.38.Aw: general properties of QCD (dynamics, confinement, etc.)
%12.38.Gc: Lattice QCD calculations
%11.10.Wx: finite-temperature field theory
%11.30.Rd: chiral symmetries

\maketitle

\section{Introduction}\label{sec:intro}
The QCD phase diagram continues to be subject of intense theoretical and experimental studies. The region of high baryon density at relatively low temperatures is of particular relevance for the inner cores of neutron stars, and at somewhat higher temperatures for neutron-star mergers. It is probed experimentally in the beam-energy scan at RHIC and the future heavy-ion programs at J-PARC, NICA and FAIR. In this regime one usually expects a chiral first-order transition ending a critical point, but inhomogeneous phases or more exotic states of matter like a quarkyonic phase have been proposed to occur as well. At even higher densities, beyond reach of current experiments and astrophysical observations,  asymptotic freedom and the attractive perturbative interactions between quarks close to the Fermi surface entail the formation of Cooper pairs and color superconductivity.  

Unfortunately, QCD at high densities remains inaccessible to stochastic integration methods, since the fermion determinant becomes complex at finite chemical potential $\mu$. This leads to an insurmountably hard fermion-sign problem precisely where a finite baryon density starts to build up in the ground state. 
The problem does not arise on the other hand in certain QCD-like theories, e.g. two-color QCD (QC$_2$D) or G$_2$-QCD \cite{Maas:2012wr,Wellegehausen:2013cya}, which show chiral symmetry breaking, confinement and asymptotic freedom as well. These theories can thus be approached with standard Monte-Carlo techniques on the lattice and provide therefore interesting testbeds to develop and test algorithms for QCD at finite density. 

Beside this technical aspect, QCD-like theories are interesting in their own right. On the lattice, QC$_2$D has been studied with staggered \cite{Kogut:2003ju,Hands:1999md,Hands:2000ei,Kogut:2001if,Kogut:2001na,Chandrasekharan:2006tz,Braguta:2016cpw,Holicki:2017psk,Bornyakov:2017txe,Astrakhantsev:2018uzd} 
and Wilson fermions \cite{Hands:2006ve,Hands:2010gd,Hands:2011ye,Cotter:2012mb,Boz:2015ppa,Itou:2018vxf}. 
In contrast to QCD, the color-singlet baryons are diquarks and hence bosonic in two-color QCD, for example, while fermionic baryons do not exist in its spectrum. The physics of the bosonic diquark baryons qualitatively resembles QCD at finite isospin density with pion condensation \cite{Son:2000xc} and is by now fairly well understood \cite{Detmold:2012wc,Kamikado:2012bt,Boettcher:2014xna}. There are firm predictions for diquark condensation when $\mu $ reaches half the pion mass $m_\pi$ from chiral effective field theory and random matrix theory \cite{Kogut:1999iv,Kogut:2000ek,Kogut:2001if,Splittorff:2001fy,Brauner:2006dv,Kanazawa:2009ks,Kanazawa:2009en,Kanazawa:2011tt}, and model studies of the BEC-BCS crossover inside the condensed phase \cite{Sun:2007fc,He:2010nb,Strodthoff:2011tz,Strodthoff:2013cua}. 
In QC$_2$D the lightest diquarks play a dual role as two-color baryons and pseudo-Goldstone bosons of the dynamical breaking of an extended chiral symmetry. When they condense, they are expected to form a superfluid which changes in nature from a Bose-Einstein condensate of tightly bound diquarks to a BCS-like pairing of quarks as chiral symmetry gets gradually restored with increasing density.

\begin{figure}[htb]
%	\centering
  \includegraphics[width=0.5\textwidth]{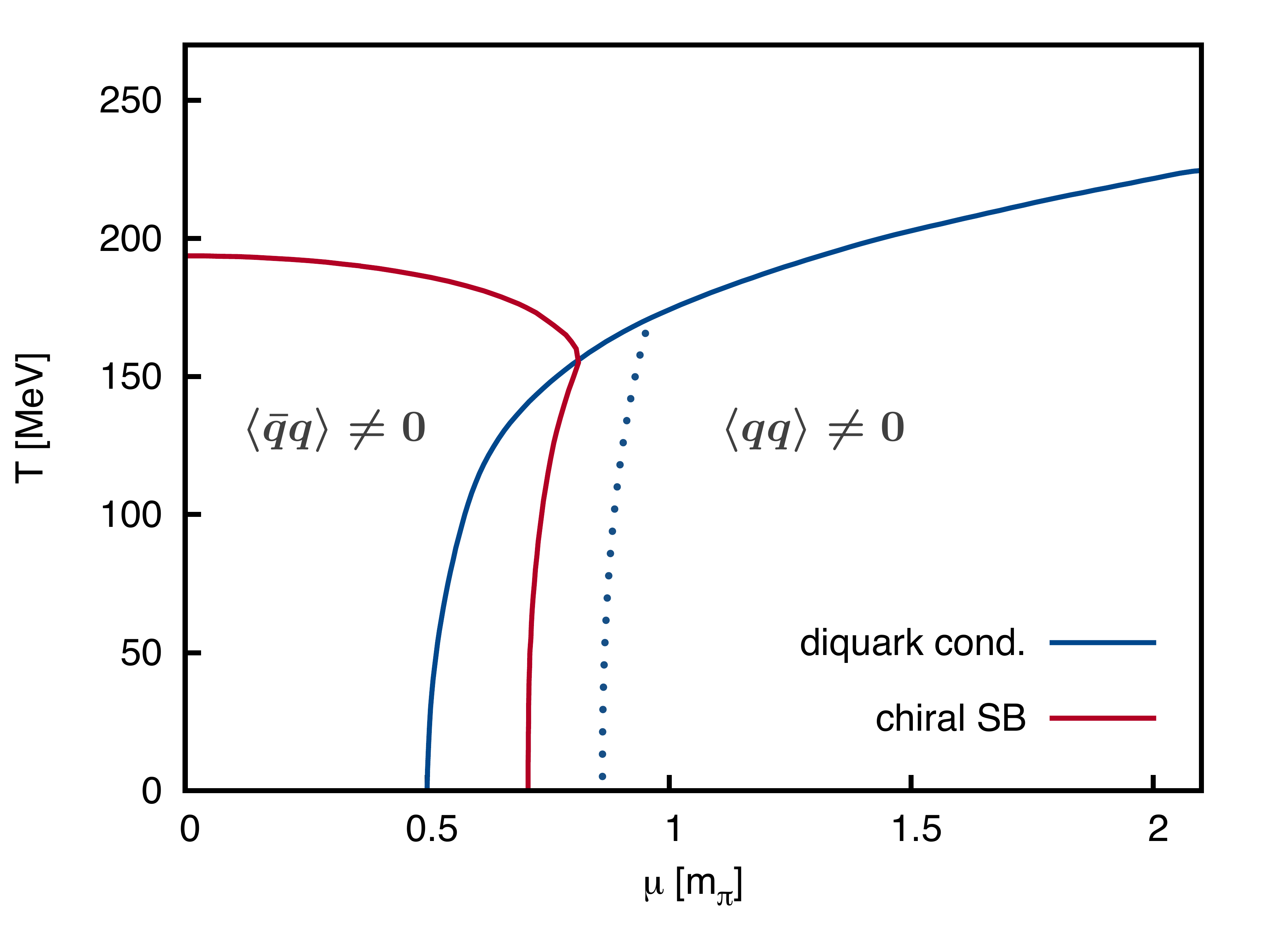}
  
        \vspace*{-.4cm}
	\caption{Phase diagram of the quark-meson-diquark model for QC$_2$D from the functional renormalization group
          \cite{Strodthoff:2011tz,Strodthoff:2013cua}:
          half-value of the chiral condensate (red), and second-order phase boundary for diquark condensation (solid blue) with rough indication of the BEC-BCS crossover (dotted blue).}
	\label{fig:qc2d_phase_diagram}
\end{figure}

The phase diagram of the quark-meson-diquark model for QC$_2$D from the functional renormalization group \cite{Strodthoff:2011tz} is shown in Figure \ref{fig:qc2d_phase_diagram}. The qualitative features resemble lattice results \cite{Cotter:2012mb,Boz:2015ppa}, especially when the Polyakov-loop variable is included in the effective model description \cite{Strodthoff:2013cua}. Evidence of the BEC-BCS crossover inside the diquark-condensation phase was also provided from lattice simulations \cite{Braguta:2016cpw} albeit still close to the bulk phase of $SU(2)$ (see below).

Anti-unitary symmetries of the Dirac operators in QCD-like theories without a fermion-sign problem are both, a blessing and a curse for the rooted-staggered fermion formulation. On one hand, phase ambiguities when rooting a complex determinant \cite{Golterman:2006rw} do not occur. In fact, even for a single staggered fermion the determinant remains positive at finite $\mu $, whereas this requires two flavors of continuum Dirac fermions. This is because of the missing $C^2 = - 1$ from the charge conjugation matrix $C$ for Dirac spinors in the anti-unitary symmetries of staggered Dirac operators. On the other hand, it implies that the corresponding Gaussian chiral random matrix ensembles get swapped, staggered fermions reflect the behavior of the Gaussian symplectic ensemble (GSE) when the continuum Dirac fermions show that of the Gaussian orthogonal ensemble (GOE) and vice versa. In particular, the staggered Dirac operator of fundamental quarks in QC$_2$D has the GSE Dyson index $\beta_D= 4$, while for continuum or Wilson fermions it is the GOE one, $\beta_D=1$. For adjoint quarks in any-color QCD (or fundamental quarks in $G_2$-QCD) it is just the other way round. This is why the sector of positive fermion determinant of QC$_2$D with adjoint quarks was studied as a replacement for the continuum theory within the correct random matrix ensemble in the early days \cite{Hands:2000ei}.

Here we have addressed the following question: with the full $SU(4)$ taste symmetry in the continuum limit, it should be possible to define a standard (tasteless) charge conjugation from that for staggered quarks \cite{Golterman:1984cy} as well. Does this imply that the correct symmetry breaking pattern, corresponding to the random matrix ensemble of the continuum two-color Dirac operator is recovered also with staggered quarks in the continuum limit? 

The answer seems to be positive. The particular evidence for this that we provide is the behavior of the Goldstone pion inside the diquark condensation phase which shows the characteristic change indicative of the change of the Dyson index as the continuum limit is approached:

In the continuum, the extended $SU(2\Nf)$ chiral symmetry is dynamically broken down to the compact symplectic  $Sp(N_\text{f})$ with fundamental quarks and Dyson index $\beta_D =1 $ \cite{Kogut:2000ek}.\footnote{For adjoint quarks in any-color QCD or fundamental quarks in $G_2$-QCD with Dyson index $\beta_D=4$ in the continuum it breaks down to $\mathrm{Spin}(2\Nf)$, the double cover of $SO(2\Nf)$.} For $\Nf = 2$ on the bosonic level this amounts to the simple vector-like breaking of $SO(6)\to SO(5)$ with coset $S^5 $ and five Goldstone bosons, three pions and a scalar (anti-)diquark pair.
The exact chiral symmetry extending the usual $U(1)_e \times U(1)_o$ of the staggered action in the two-color case is $U(2)$ on the other hand \cite{Hands:1999md}. With fundamental quarks it breaks down to $U(1)_V$. Up to the Goldstone pion of the broken $U(1)_\epsilon$ it therefore resembles the $\Nf =1$ case of continuum quarks with Dyson index $\beta_D = 4$ as in adjoint QCD \cite{Kogut:2000ek} or $G_2$-QCD \cite{Wellegehausen:2015iea} which is $SU(2) \to U(1)$.\footnote{For adjoint staggered quarks in QC$_2$D one has $U(2) \to Sp(1) \sim SU(2) $ and hence only the Goldsone pion of the broken $U(1)_\epsilon$ corresponding to having no Goldstone bosons with Dyson index $\beta_D = 1$ for $\Nf =1$ in the continuum theory.}  

The effective field theory prediction for the Goldstone spectrum with the $U(2)\to U(1) $ chiral symmetry breaking of the fundamental staggered  two-color action was explicitly worked out in \cite{Kogut:2003ju}. Most important for our purposes is the behavior of the Goldstone pion inside the diquark condensation phase at $\mu \ge m_\pi/2$ where it resembles that of the symmetric pion branch $P_S$ of the Dyson index $\beta_D = 4$ case in the continuum \cite{Kogut:2000ek}, with a mass that decreases $\sim m_\pi^2/2\mu $, although this branch strictly speaking only exist for $\Nf \ge 2 $ there.

We will demonstrate that the behavior of this Goldstone pion branch in the continuum limit indeed changes to that of the symmetric $P_S$ mode in the $\beta_D  = 1$ case of the continuum theory which increases $\sim 2\mu$ inside the diquark condensation phase \cite{Kogut:2000ek}. Again this mode exists in the continuum only for $\Nf \ge 2 $, starting with a multiplicity of 3 for the three degenerate pions in the two-flavor case. This again indicates that the taste symmetry needs to be at least partially restored to achieve this. 

Here it is also important to note that the previous results with staggered
quarks in QC$_2$D that led to the $\beta_D = 4$ Goldstone spectrum were obtained inside a so-called \textit{bulk phase} on the lattice, where the lattice spacing is almost independent of the inverse gauge coupling. The large effects of this on spectroscopy and thermodynamics at vanishing chemical potential have been investigated in Ref.~\citep{SchefflerDiss2015}. The influence of bulk effects on simulations at finite density has not yet been discussed and is thus the main focus of our study.
Indeed, we will find that inside the bulk phase, the Goldstone spectrum for Dyson index $\beta_D = 4$ as for adjoint continuum quarks is reproduced, while outside the bulk phase we observe the correct Goldstone spectrum of two-color QCD, corresponding to the Dyson index $\beta_D = 1$ of the continuum Dirac operator, also with rooted-staggered quarks. 
We demonstrate that this change of Dyson index is reflected in the eigenvalue statistics of the Dirac operator. By computing the unfolded level
spacings, we demonstrate that the distribution of eigenmodes is completely dominated by the GSE in the bulk phase, but obtains larger contributions
from the GOE as one leaves the bulk phase. This change turns out to be continuous, and builds up starting with the low-lying eigenmodes. 
As infrared physics is controlled by these lower levels, we observe that one correctly reproduces the continuum theory Goldstone spectrum even when
the higher eigenmodes are still dominated by the GSE. 

This paper is organized as follows: In Section~\ref{sec:latsetup} we introduce the lattice action, including a diquark source term, and 
the simulation parameters used in this work. In Section \ref{sec:xpt} we study the quark-number density and chiral and diquark condensates 
for small values of the gauge coupling (which turns out to be inside the bulk phase) and compare our results to the predictions of leading 
order chiral perturbation theory. Section~\ref{subsec:bulkphase} explains the bulk phase and 
introduces its order parameter, the ${Z}_2$-monopole density, ending with a discussion on 
new lattice parameters to suppress bulk effects. In Section~\ref{sec:latsat} we then repeat the study of Section \ref{sec:xpt} 
with our new set of parameters and discuss lattice discretization and finite volume effects, such as the effective quenching of 
the theory in the saturated regime and problems with additive renormalization of the chiral condensate at finite $\mu$. 
In Section~\ref{sec:patsymbreak} after providing a detailed description of chiral symmetry-breaking pattern of staggered fermions,
we present our main result, which is the numerical measurement and comparison of the Goldstone spectrum at non-vanishing chemical potential, 
inside and outside the bulk phase. 
And finally, in Section~\ref{sec:levelspac} we present the unfolded level spacings of the Dirac operator and demonstrate that change of 
Goldstone spectrum is accompanied by a change of eigenvalue distribution. We end with our conclusion and an outlook in Section~\ref{sec:concandoutl}.

\section{Lattice Setup}\label{sec:latsetup}
For the low-temperature scan of the Goldstone spectrum of two-color QCD, we use standard rooted staggered fermions, with the Dirac operator
\begin{equation}
	\begin{split}
		D_{x,y}(\mu) =a m \delta_{x,y} + \sum_{\nu=1}^4 &\frac{\eta_\nu(x)}{2a} \left( e^{a\mu\delta_{\nu,4}} U_\nu(x) \delta_{x+\hat\nu,y}\right.  \\
		&\left.- e^{-a\mu\delta_{\nu,4}} U^\dag_\nu(x-\hat\nu) \delta_{x-\hat\nu,y}  \right)
	\end{split}
\end{equation}
at non-vanishing baryon chemical potential $\mu$. To study spontaneous
symmetry breaking and competing order on a finite lattice, we add a
diquark source $\lambda$ corresponding to a Majorana mass term in the
Lagrangian, which explicitly breaks the chiral $U(2)$ symmetry of the
massless staggered action down to a $U(1)$ as well but in a
direction different from that of the Dirac mass term
\cite{Hands:1999md,Hands:2000ei,Kogut:2001if,Kogut:2001na}, 
\begin{equation}
	S_\text{f} = \bar{\chi} \, D(\mu)\, \chi + \frac \lambda 2 \left( \chi^T\,\tau_2\,\chi+\bar{\chi}\,\tau_2\,\bar{\chi}^T \right).
	\label{eq:diquark_action}
\end{equation}
Physical results are then retrieved in the $\lambda\rightarrow 0$ limit. The diquark condensate is obtained from
\begin{equation}
	\langle q q \rangle = \langle \chi^T \tau_2 \chi \rangle \propto \left.\frac{\partial \ln Z}{\partial \lambda} \right|_{\lambda\rightarrow0} .
\end{equation}
The staggered fermion action is conveniently expressed in a Nambu-Gorkov basis
%\text{$\Psi = \begin{pmatrix} \tau_{2} \overline \psi^T \\ \psi \end{pmatrix}$}, \text{$\Psi^T = \left( \overline\psi, \psi^T \right)$},
\begin{equation}
	S_\text{f} = \frac 1 2 \left( \bar{\chi}, \chi^T\tau_2 \right) \,A\, \begin{pmatrix} \tau_{2} \bar{\chi}^T \\ \chi \end{pmatrix}\,,
	\label{eq:gorkovaction}
\end{equation}
with an inverse Nambu-Gorkov propagator
\begin{equation}
 A=\begin{pmatrix} \lambda  & D(\mu) \\ -D^\dagger(\mu) & \lambda  \end{pmatrix}.
 \label{eq:gorkovinvprop}
\end{equation}
Grassmann integration produces the square root of the determinant of $A$ in the path integral measure. This is seen most easily when considering $(\chi, \bar \chi^T)$ as a set of independent real Grassmann variables, whose Gaussian integral results in a (positive) Pfaffian which agrees with $\sqrt{\det A}$, where
\begin{equation}
  \det A = \det \left( D^\dag(\mu)D(\mu) + \lambda^2 \right).
  \label{eq:detA}
\end{equation}
In fact, a direct use of hybrid Monte-Carlo (HMC) based on a Gaussian pseudo-fermion integral over $(AA^\dagger)^{-1}$ would even produce $(\det A)^2$ in the measure. However, because
\begin{equation*}
  A^\dagger A = AA^\dagger = \begin{pmatrix}
    D(\mu)D(\mu)^\dag + \lambda^2 & 0 \\
    0 & D^\dag(\mu)D(\mu) + \lambda^2
  \end{pmatrix}
\end{equation*}
is block diagonal with both blocks having the same determinant, we can use size-half pseudo-fermion fields (in the Nambu-Gorkov space) to remove this further doubling. On the other hand, we can not use size-half fields in the even-odd staggered lattice at finite $\mu$. This means that without any rooting, we compute $\det A$ and describe eight fermion species instead of the usual four staggered tastes.
Therefore we use standard rooting techniques to compute $(\det A)^{N_\text{f}/8}$ for each continuum flavor, i.e.~we take the fourth root to simulate with $N_\text{f}=2$ as in Ref.~\cite{Braguta:2016cpw}. 
The rooting is achieved by a rational approximation of the fermion matrix in the pseudofermion action of the HMC algorithm.

The diquark source explicitly breaks baryon number conservation, or more precisely the $U(1)_V$ of the staggered action, and hence in addition to the usual $\left\langle \psi\bar{\psi} \right\rangle$ contractions in the calculation of the correlation functions, we now also have $\left\langle \psi\psi \right\rangle$ and $\left\langle \bar{\psi}\bar{\psi} \right\rangle$ contractions.
These correspond to the diagonal terms of the propagator $G=A^{-1}$ obtained from Eq.~\eqref{eq:gorkovinvprop},
\begin{equation}
G = \left( \begin{array}{cc}
\left(DD^{\dagger}+\lambda^{2}\right)^{-1}\lambda	&-\left(DD^{\dagger}+\lambda^{2}\right)^{-1}D\\
\left(D^{\dagger}D+\lambda^{2}\right)^{-1}D^{\dagger}	&\left(D^{\dagger}D+\lambda^{2}\right)^{-1}\lambda\\
\end{array} \right)\ .
\label{eq:full_prop}
\end{equation}

In this paper we compare results with $N_\text{f}=2$ staggered flavors at $\beta=1.5$ on a $12^3\times24$ lattice with $am=0.025$, which turns out to
be deep inside the bulk phase, to $N_\text{f}=2$ staggered flavors at $\beta=1.7$ on a  $16^3\times 32$ lattice with $am=0.01$ and an improved gauge action so that this is just outside the bulk phase. The parameters of the simulations at $\beta = 1.5$ with the unimproved gauge action correspond to those used in Ref.~\cite{Kogut:2003ju} for $\Nf=4$.  

\section{Effective Field Theory Predictions}\label{sec:xpt}
Kogut et al.~have studied the symmetries of QCD-like theories at
finite baryon density with pseudoreal quarks in the fundamental
representation using chiral effective Lagrangians
\cite{Kogut:2000ek,Kogut:2003ju}. A linear sigma model for the
symmetry-breaking pattern and Goldstone spectrum of the staggered
two-color action was used to describe the data in
\cite{Kogut:2003ju}. For the purpose of illustrating the basic
features of the diquark-condensation transition at $\mu = \mu_c =
m_\pi/2$ here we fit our data to the somewhat simpler form of the
leading-order chiral perturbation theory ($\chi$PT) predictions from
the non-linear sigma model \cite{Kogut:2000ek}. This describes the
rotation of the vacuum alignment from the chiral $\langle \bar q q
\rangle$ into the diquark $\langle  q q \rangle $ condensate at a
fixed   
\begin{equation}
	\Sigma_\text{c} = \sqrt{\langle\bar{q}q\rangle^2+\langle q q\rangle^2} \equiv 2 \, N_\text{f}\,G\, .
\end{equation}
With explicit diquark source  $\lambda $, the rotation
angle $\alpha(\mu)$ is obtained from 
\begin{equation}
  \mu^2 \cos\alpha \sin\alpha = \mu_c^2 \big( \sin\alpha - \frac{\lambda}{m} \cos\alpha \big) \, ,
  \label{eq:xpt_angle}
\end{equation}
such that a non-zero value $\alpha_0$ is obtained already at $\mu = 0$
which depends on the relative size of the Majorana and Dirac quark
masses, i.e.~$\tan\alpha_0 =\lambda/m$. Chiral and diquark condensate, and 
quark-number density $n$ as functions of $\mu$  are then given by
\begin{equation}
	\begin{aligned}
		\langle \bar{q} q \rangle  &= 2 N_\text{f}\, G \cos \alpha\,, \\
		\langle q q \rangle &= 2 N_\text{f}\, G \sin \alpha\,, \\
		n &= 8 N_\text{f}\, F^2 \mu \sin^2 \alpha\, . 
	\end{aligned}
	\label{eq:xpt_condensates}
\end{equation}

As a first test we have performed simulations with the parameters of
Ref.~\cite{Kogut:2003ju}, i.e.~a lattice gauge coupling of
$\beta=1.5$, quark mass $a m=0.025$, and diquark source $a
\lambda=0.0025 $ on a $12^3 \times 24$ lattice and the standard
 Wilson plaquette action, so as  to reproduce their results with the
 square root of the determinant in (\ref{eq:detA}) for $\Nf = 4$.

 \begin{figure}[htb]
	\centering
	\includegraphics[width=0.95\linewidth]{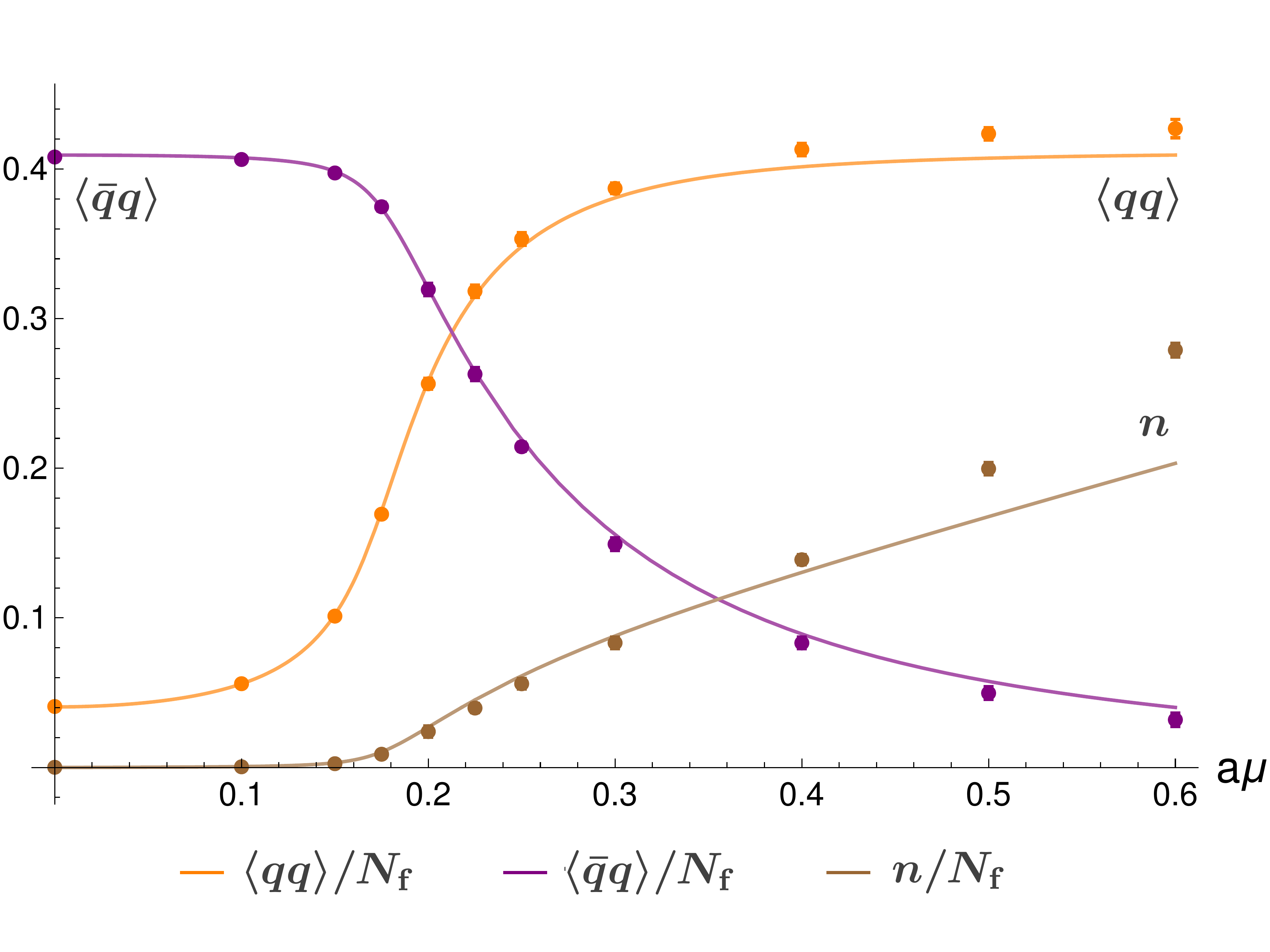}
	\caption{Fit of lattice data to leading-order $\chi$PT form of
          chiral (purple)  and diquark (orange) condensates,
          and quark-number density (brown)
          from Eqs.~(\ref{eq:xpt_angle}) and (\ref{eq:xpt_condensates}),
          with $N_\text{f}=2$, $\beta = 1.5$, $a m=0.025$, $a\lambda=0.0025$
          on a $12^{3}\times 24$ lattice with unimproved Wilson gauge
          action.} 
	\label{fig:xpt_fit}
\end{figure}

 With the fourth root for $N_\text{f}=2$ continuum flavors and the same lattice
 parameters we have then obtained the results shown in
 Figure~\ref{fig:xpt_fit}. They are fitted to the
 leading-order forms from chiral perturbation theory in
 Eq.~\eqref{eq:xpt_condensates} with \eqref{eq:xpt_angle} for the
 vacuum alignment angle $\alpha(\mu)$. From these fits we obtain the
 fit parameters $F,G$, and the critical chemical potential, which
 results as $ a\mu_\text{c} = 0.1889(5)$.  Within the errors this agrees 
 with our spectroscopic result from the pion correlator at $\lambda= 0.0025, \mu=0$
 which yields $a m_\pi/2=0.1887(6)$. The agreement between fits and
 data in Figure \ref{fig:xpt_fit} is nearly perfect up to $a \mu\sim 0.3$. 

The deviations at larger chemical potentials were attributed to a
$\mu$-dependence of the total condensate $\Sigma_\text{c}$ in
Ref.~\cite{Kogut:2003ju} where linear sigma model fits were therefore
used instead. These imply $\langle qq\rangle \sim \mu$ and hence $n
\sim \mu^3$ at large $\mu$, as predicted for a BCS-like pairing in QCD at
large isospin density already in \cite{Son:2000xc}. The corresponding
rise in the isospin density beyond the $\chi $PT prediction was
observed in lattice QCD simulations \cite{Detmold:2012wc}, and it was
traced to the BEC-BCS crossover in a functional renormalization
group study of the quark-meson model as an effective theory with linearly
realized chiral symmetry and order-parameter fluctuations
beyond mean field \cite{Kamikado:2012bt}.  

The same interpretation of this rise in the diquark density, beyond
the $\chi$PT prediction, as an indication of the BEC-BCS crossover in
QC$_2$D, was also adopted in Ref.~\cite{Braguta:2016cpw}. Because 
these results were obtained on rather coarse lattices, it is therefore
important to verify that they are not qualitatively affected by strong
discretization artifacts such as the $Z_2$ monopoles in the bulk phase
of $SU(2)$.

\section{The Bulk Transition}\label{subsec:bulkphase}

Most lattice gauge theories, as for instance $SU(2)$, $SU(3)$ or $G_2$ gauge theory, exhibit a bulk phase in the 
strong-coupling regime, characterized by the presence of unphysical lattice artifacts such as electric vortices and 
magnetic monopoles \cite{Bhanot:1981eb, Halliday:1981te, Halliday:1981tm, Mack1982, Barresi:2003yb, Heller:1995bz}. 
These dominate the ultraviolet behaviour, such that the lattice spacing is nearly independent
of the coupling constant, and taking a continuum limit is not possible.  
In the physical weak-coupling regime the short distance physics is governed by asymptotic freedom and the continuum limit 
is approached by $\beta \to \infty$. Depending on which gauge group and which representation, 
both regions are either separated by a true phase boundary or a cross-over, the later being the case for the fundamental 
representation of $SU(2)$ considered here (cf. Figure $1$ in Ref. \cite{Barresi:2003yb}).
The bulk transition is almost independent of the lattice size and persists also in the presence of fermions \cite{Aoki:2004iq,Brown:1992fz}.

An order parameter for the strong-coupling to weak-coupling transition is the $Z_2$ monopole density \cite{Halliday:1981tm} 
\begin{equation}
	 \langle z \rangle = 1 - \frac{1}{N_C}  \sum_{C} \prod_{P\in\partial C} \sgn \tr P,
\end{equation}
where $\sum_C$ runs over all elementary cubes of the lattice.
$\langle z \rangle$ is sensitive to preferred signs of the plaquettes on the faces of these cubes, which are aligned below the bulk transition:
In the bulk phase $\langle z \rangle$ is non-vanishing, while it vanishes in the physical weak-coupling regime.

\begin{figure}[htb]
	\centering
	\includegraphics[width=0.5\textwidth]{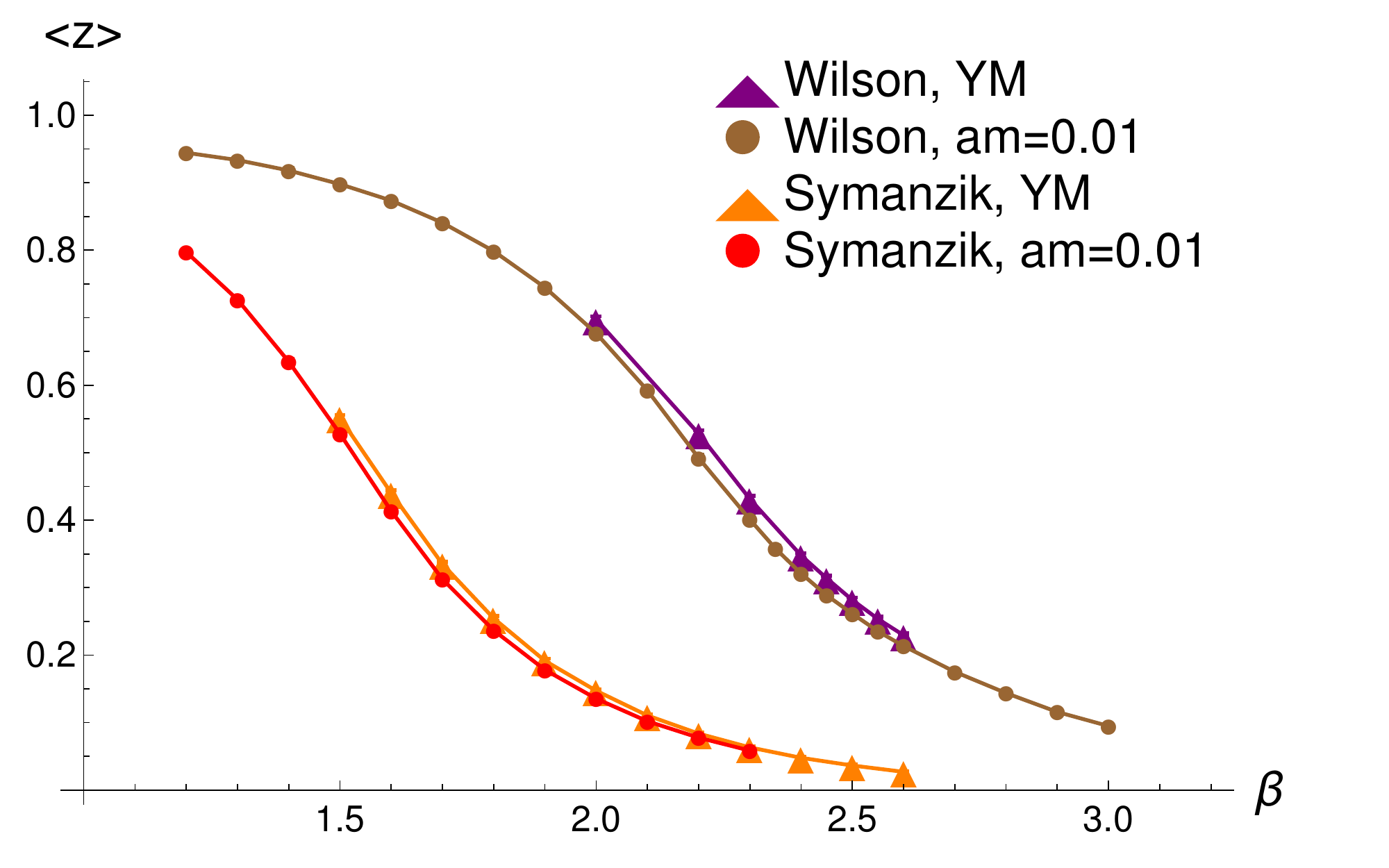}
	\caption{Comparison of the $Z_2$ monopole density on a $N_s=12$, $N_t=8$ lattice for Wilson and tree-level improved 
Symanzik gauge action in pure Yang-Mills theory and with standard staggered fermions for $a m=0.01$ as a function of the inverse gauge coupling $\beta$.}
	\label{fig:bulk_phase_dependencies}
\end{figure}

With $\beta=1.5$ and the unimproved Wilson gauge action, we have found the $Z_2$ monopole density 
at $a\mu=0$ to be $\langle z \rangle = 0.884040(95)$, and thus we expect bulk effects to be dominant in this regime.
We have also confirmed that $\langle z \rangle$ is only very weakly affected by the inclusion of dynamical quarks.
For a fixed inverse gauge coupling $\beta$, the monopole density can be significantly reduced with Symanzik's gauge action 
\cite{Weisz:1983bn} (we employ the tree-level improved variant here). 
This is illustrated in Figure \ref{fig:bulk_phase_dependencies}, which shows the $\beta$-dependence 
of $\langle z \rangle$ for the improved and unimproved actions, with and without dynamical fermions in each case.

\begin{figure}[htb]
	\centering
	\includegraphics[width=0.5\textwidth]{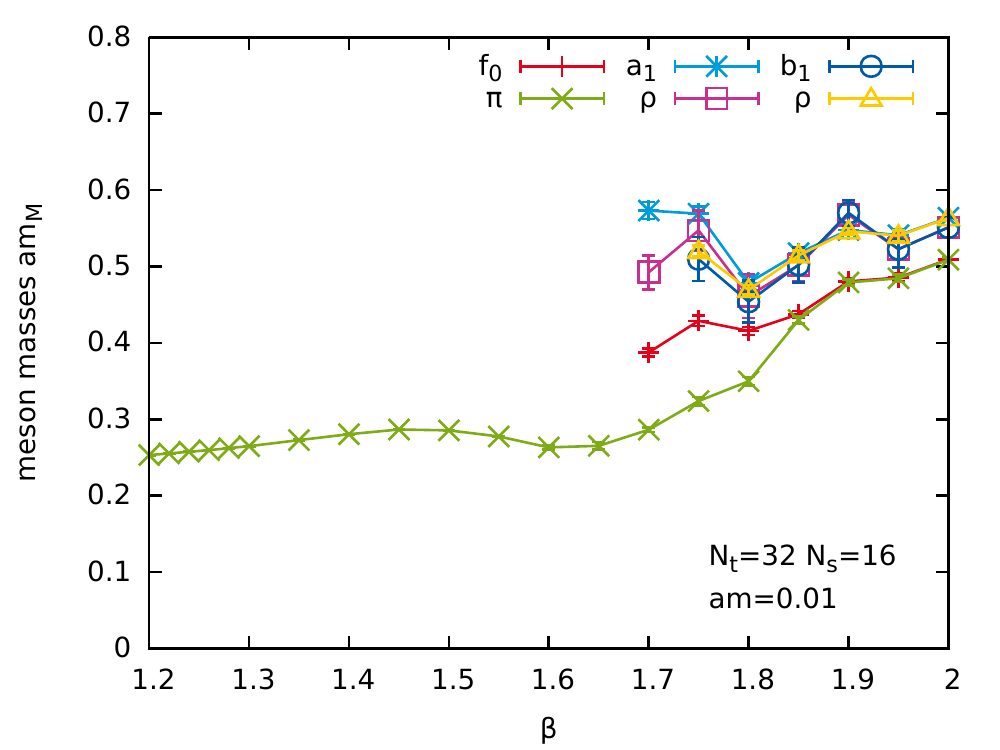}
	\caption{$\beta$-dependence of meson masses, obtained with standard staggered fermions and tree-level improved Symanzik
gauge action. Figure taken from \cite{SchefflerDiss2015}.}
	\label{fig:meson_masses_over_beta}
\end{figure}

In principle, one would like to suppress $Z_2$ monopoles as much as possible.
In practice, finite volume effects become increasingly severe at larger $\beta$ due to a smaller physical 
lattice spacing $a$, and one is forced to make a compromise. This is illustrated in Figure \ref{fig:meson_masses_over_beta},
which shows the $\beta$-dependence of various meson masses obtained in a previous study 
\cite{SchefflerDiss2015}.
For small physical volumes $(aN)^4$ (larger $\beta$) the meson masses degenerate, signaling an explicit breaking
of chiral-symmetry by the finite system size, while for small inverse couplings $\beta$ bulk effects are dominant.
For our simulations of the continuum physics we therefore chose $\beta=1.7$ with $N_s=16$ and $N_t=32$, as a compromise between small 
$a$ and small $\beta$, where $m_\pi/m_\rho=0.58(5)$. Using the improved action the $Z_2$ monopole density at $\beta=1.7$ and $a\mu=0$ is 
$\langle z \rangle = 0.27340(66)$. Although there is still a substantial amount of monopoles on the lattice, 
this choice of parameters pushes the simulations on the weak-coupling 
side of the bulk crossover, as our spectroscopic results discussed below clearly demonstrate.

\section{Leaving the Bulk Phase}\label{sec:latsat}

\begin{figure}
	\centering
	\includegraphics[width=0.5\textwidth]{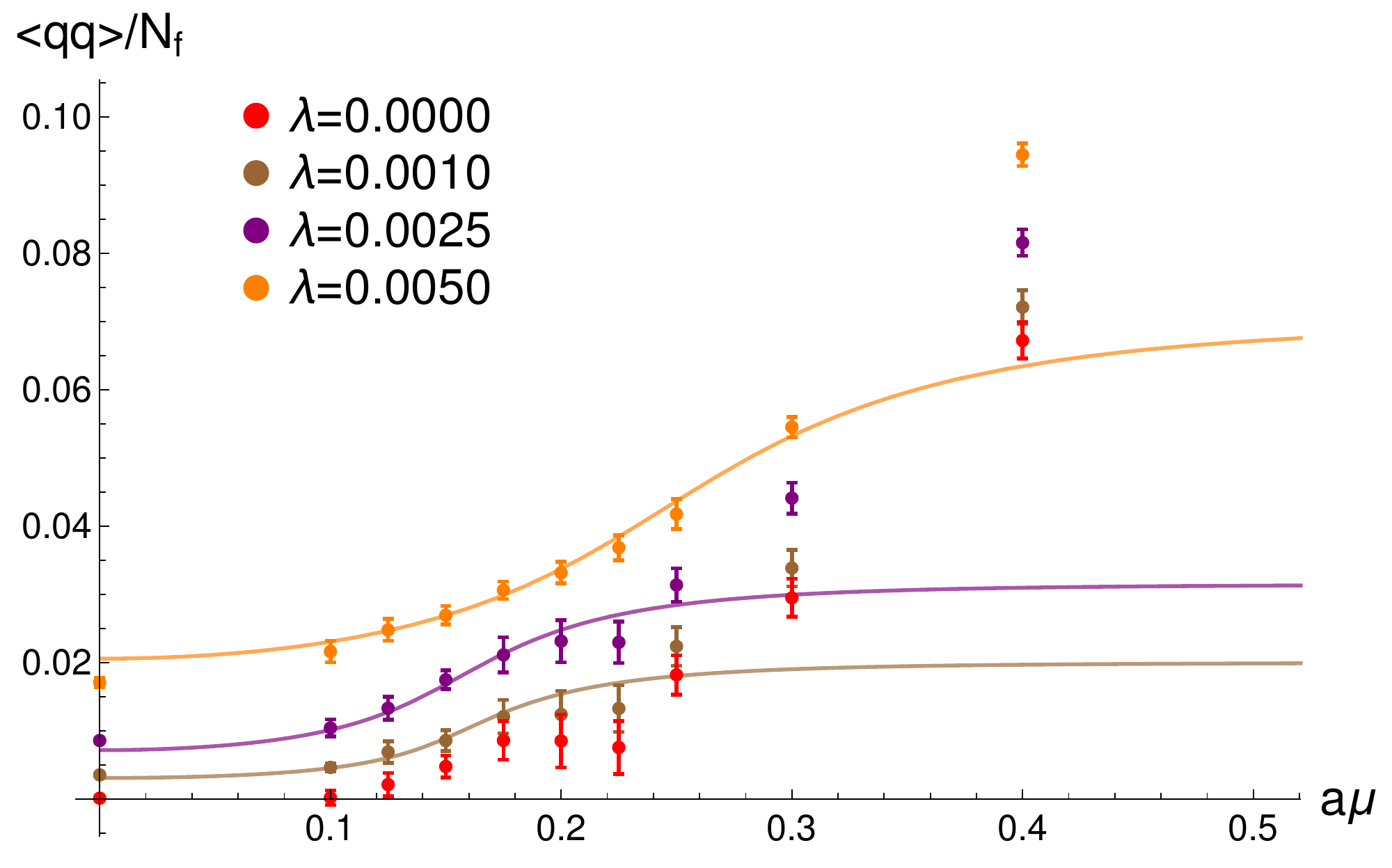}
	\includegraphics[width=0.5\textwidth]{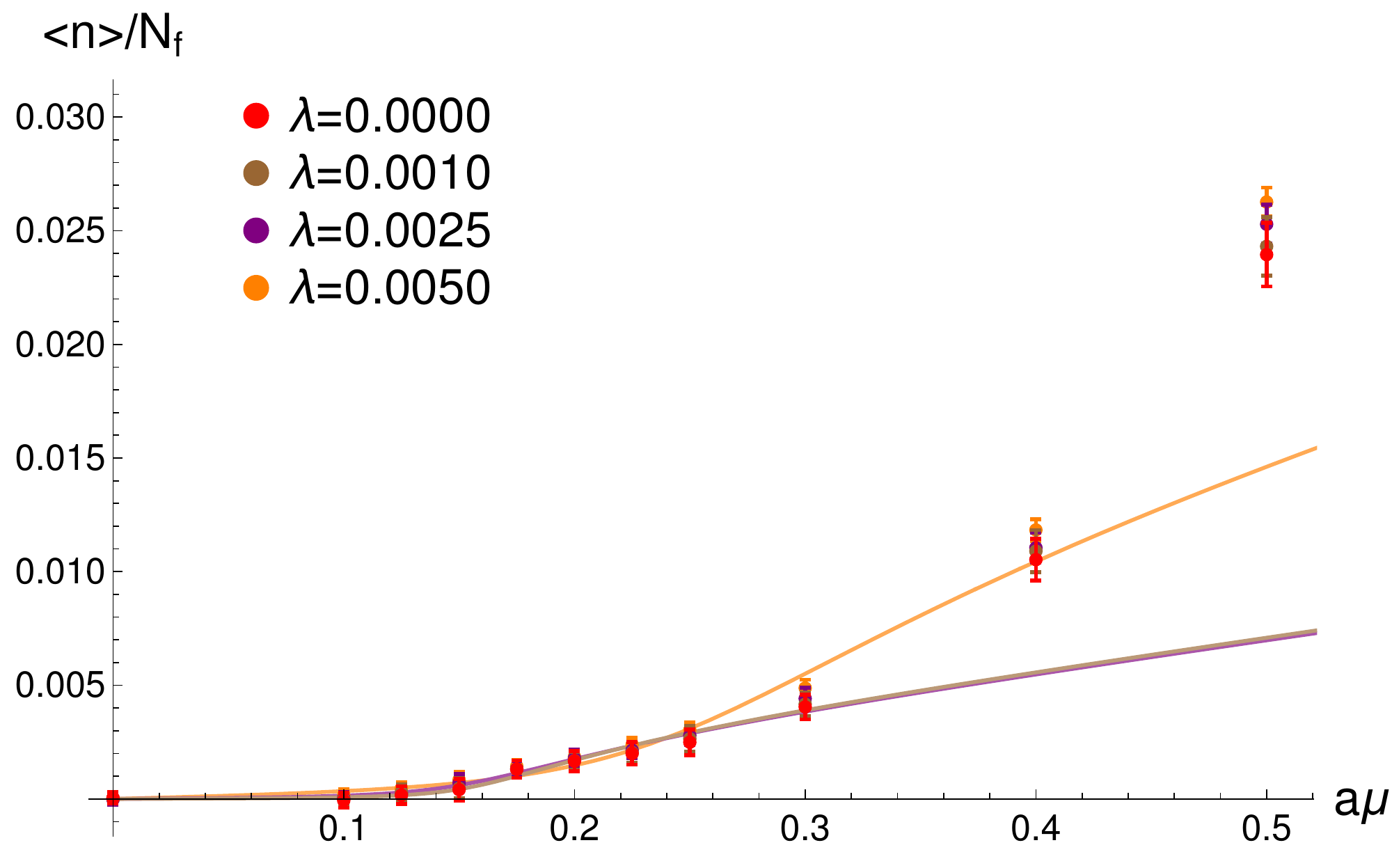}
	\caption{Fit of lattice data to leading-order $\chi$PT form of
          diquark condensate (top) and quark-number density (bottom)
          from Eqs.~(\ref{eq:xpt_angle}) and (\ref{eq:xpt_condensates}),
          with $N_\text{f}=2$, $\beta = 1.7$, $a m=0.01$, $a\lambda=0.0050;0.0025;0.0010$
 	  on a $16^{3}\times 32$ lattice with tree-level improved Symanzik gauge
          action. Points for $a\lambda=0.0$ obtained by extrapolation.}
	\label{fig:chiPT_lambda_dens_dqc}
\end{figure}

Simulating at finite $\mu$ with the improved action and our new choice of lattice parameters ($\beta=1.7$, $am=0.01$, $16^3\times 32$), 
we carry out a study of the $\mu$ dependence of different observables outside of the bulk phase.
We first observe that we can again fit the quark-number density and diquark condensate to leading-order $\chi$PT predictions 
from Eqs.~(\ref{eq:xpt_angle}) and (\ref{eq:xpt_condensates}) 
(note here that the same expressions are predicted by $\chi$PT for both the Gaussian orthogonal and Gaussian symplectic
ensembles \cite{Kogut:2000ek,Kogut:2003ju}). We simulate with three different values of the explicit
diquark source $a\lambda=0.0050;0.0025;0.0010$ and apply the fits directly at finite $\lambda$. Results are shown
in Fig. \ref{fig:chiPT_lambda_dens_dqc}, together with extrapolations to $\lambda\to 0$. 

We find that 
our results for $a\lambda=0.0010$ already agree within one standard deviation with the limit of vanishing
diquark source. Attempting to extract the critical chemical potential for diquark condensation 
from fits to $a\lambda=0.0010$ yields $a\mu_\text{c} = 0.172(21)$ however, which 
is slightly overestimated compared to our spectroscopic result ($a\mu_\text{c} = 0.1456(28)$) from 
the pion correlator at $a\lambda=0.0010,\ \mu=0$. We find that a consistent value ($a\mu_\text{c} = 0.1356(86)$)
is obtained from a $\chi$PT fit to the $\lambda\to 0$ extrapolation of $\langle qq \rangle$.
We also observe significant deviations from the $\chi$PT predictions at around $a\mu \sim 0.3$, which we interpret
as signaling the onset of the BEC-BCS crossover. We thus conclude that the behavior
of $\langle  q q \rangle $ and $\langle n \rangle$ is not qualitatively different outside of the bulk phase. 

\begin{figure}[!ht]
	\centering
	\includegraphics[width=0.5\textwidth]{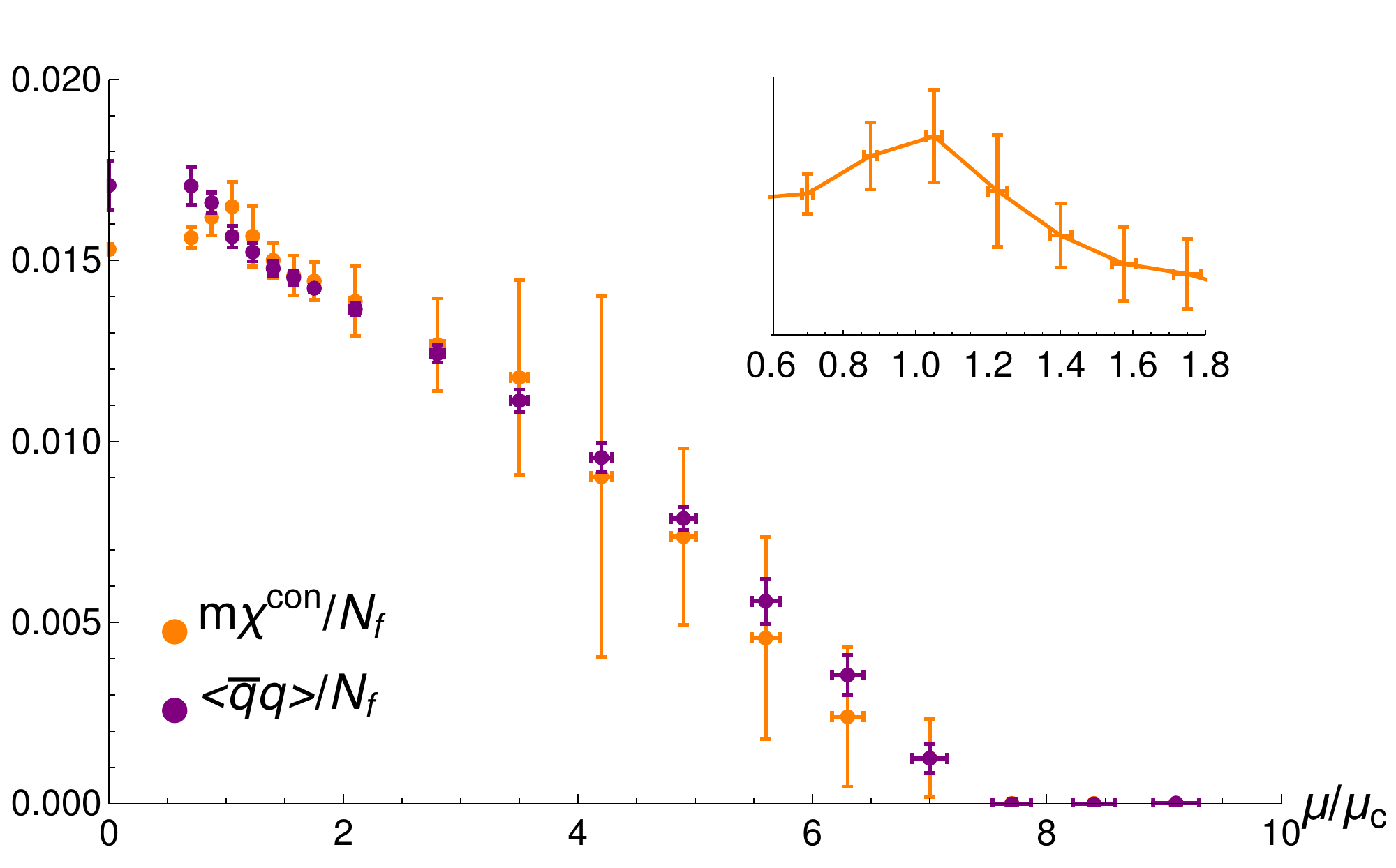}
	\caption{Chiral condensate and connected susceptibility, $16^3\times32$, \text{$a m=0.01$}, $\beta=1.7$, $a \lambda=0.005$ with a zoom to the singular contribution to $\chi^\textmd{con}$.}
	\label{fig:chiral_condensate_renorm}
\end{figure}

On the other hand, from an observed decrease of the chiral condensate above $\mu_\text{c}$ (shown in Figure \ref{fig:chiral_condensate_renorm}), 
which we did not observe at $\beta=1.5$, we infer the presence of UV-divergence, such that renormalization is required. As discussed in Ref. \cite{Unger2010}, it is 
possible to renormalize the chiral condensate at finite temperature using the chiral susceptibility $\chi_{m_q}$, as both contain
the same UV-divergent term $c_{UV}$, viz.
\begin{equation}
	\begin{split}
		\langle\bar{q} q\rangle_{m_q} &= \langle\bar{q} q\rangle_0 + c_2 m_q + \frac{c_{UV}}{a^2} m_q + \mathcal O(m_q^2) ,\\
		\chi_{m_q} &= c_2 + \frac{c_{UV}}{a^2} + \mathcal O(m_q^2).
	\end{split}
\end{equation}
Since the UV divergence
originates mainly from the connected chiral susceptibility $\chi^\textmd{con}$ (also shown in Figure \ref{fig:chiral_condensate_renorm}, we neglect $\lambda$-dependent contributions here),
a renormalized condensate can be defined as \text{$\Sigma = \langle\bar{q} q \rangle_{m_q} - m_q \chi^\textmd{con}$}. 
We observe that both the condensate and $\chi^\textmd{con}$ exhibit a similar decrease at large $\mu$
and thus conclude that the UV-divergence $c_{UV}$ is $\mu$-dependent.

\begin{figure}[!ht]
	\centering
	\includegraphics[width=0.5\textwidth]{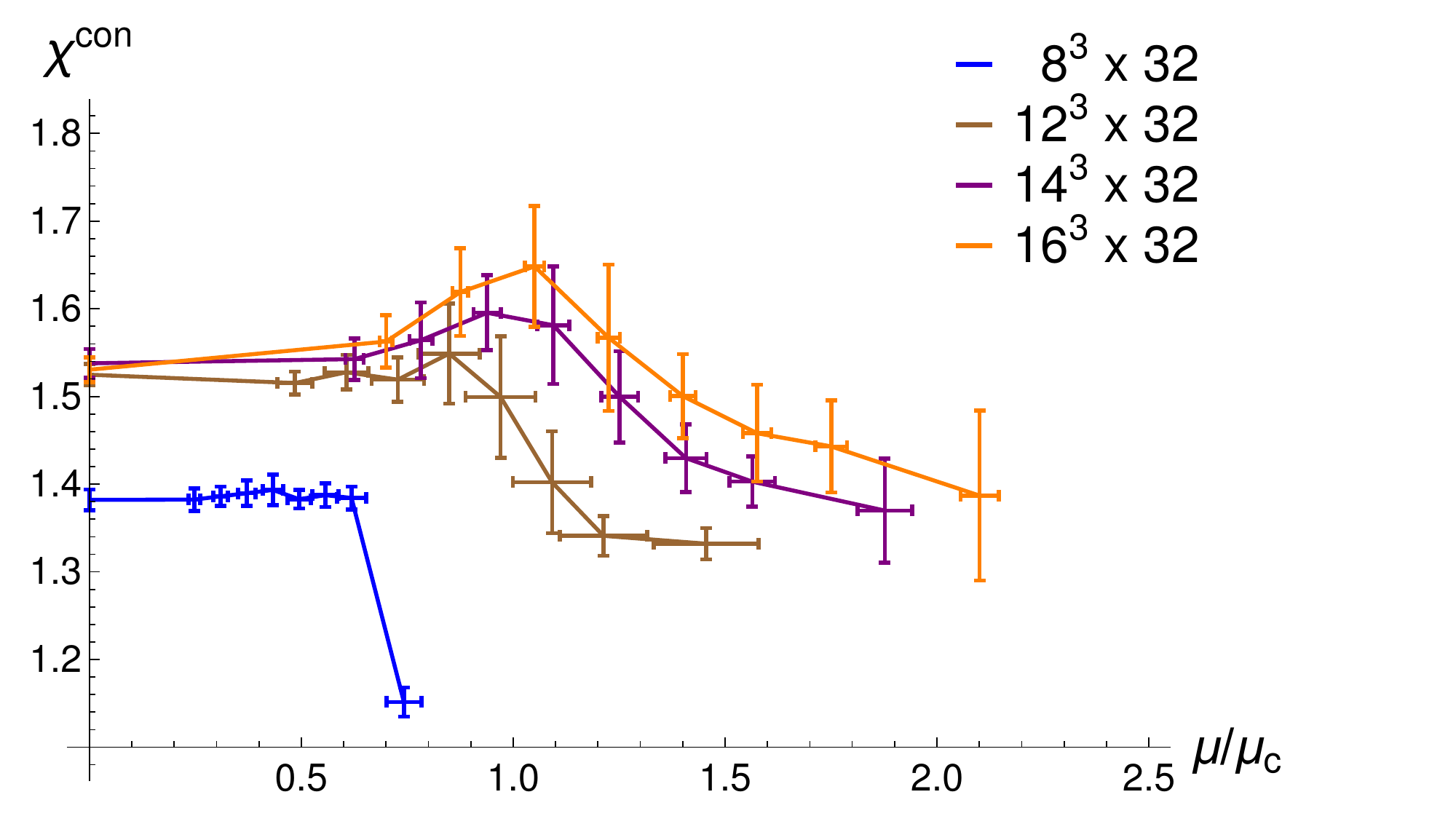}
	\caption{Volume dependence of the connected chiral susceptibility for $16^3\times32$, \text{$a m=0.01$}, $\beta=1.7$, $a\lambda=0.005$.}
	\label{fig:chiral_con_susc_vol_dep}
\end{figure}

At the chiral transition the disconnected susceptibility $\chi^\textmd{dis}$ contains a singular contribution. At the diquark condensation transition, we find that $\chi^\textmd{con}$ has a singular part as well, as its peak height is bounded from above by a 
finite volume (see Figure \ref{fig:chiral_con_susc_vol_dep}).
This singularity will dominate over $c_{UV}/a$ at finite $a$ in the infinite volume limit. 
The chiral condensate on the other hand, at zero temperature, must remain independent of $\mu $ for $\mu<\mu_c$.  
It does not have such a singular contribution and it would be unphysical to introduce one with the connected susceptibility 
subtraction.  At any rate, this would introduce a $\mu$-dependence below $\mu_c$ and hence a Silver-Blaze problem. 
Therefore, a different ($\mu$-dependent) subtraction of the chiral condensate is required.
Likewise, it is impossible to remove the UV-divergence by subtracting a heavy quark condensate like 
$\langle \bar{q} q \rangle_{m_q} - \frac{m_q}{m_q'} \langle \bar{q} q \rangle_{m_q'}$,
since the pion mass and thus the position of the diquark onset strongly depend on the quark mass.

\begin{figure}
	\centering
	\includegraphics[width=0.5\textwidth]{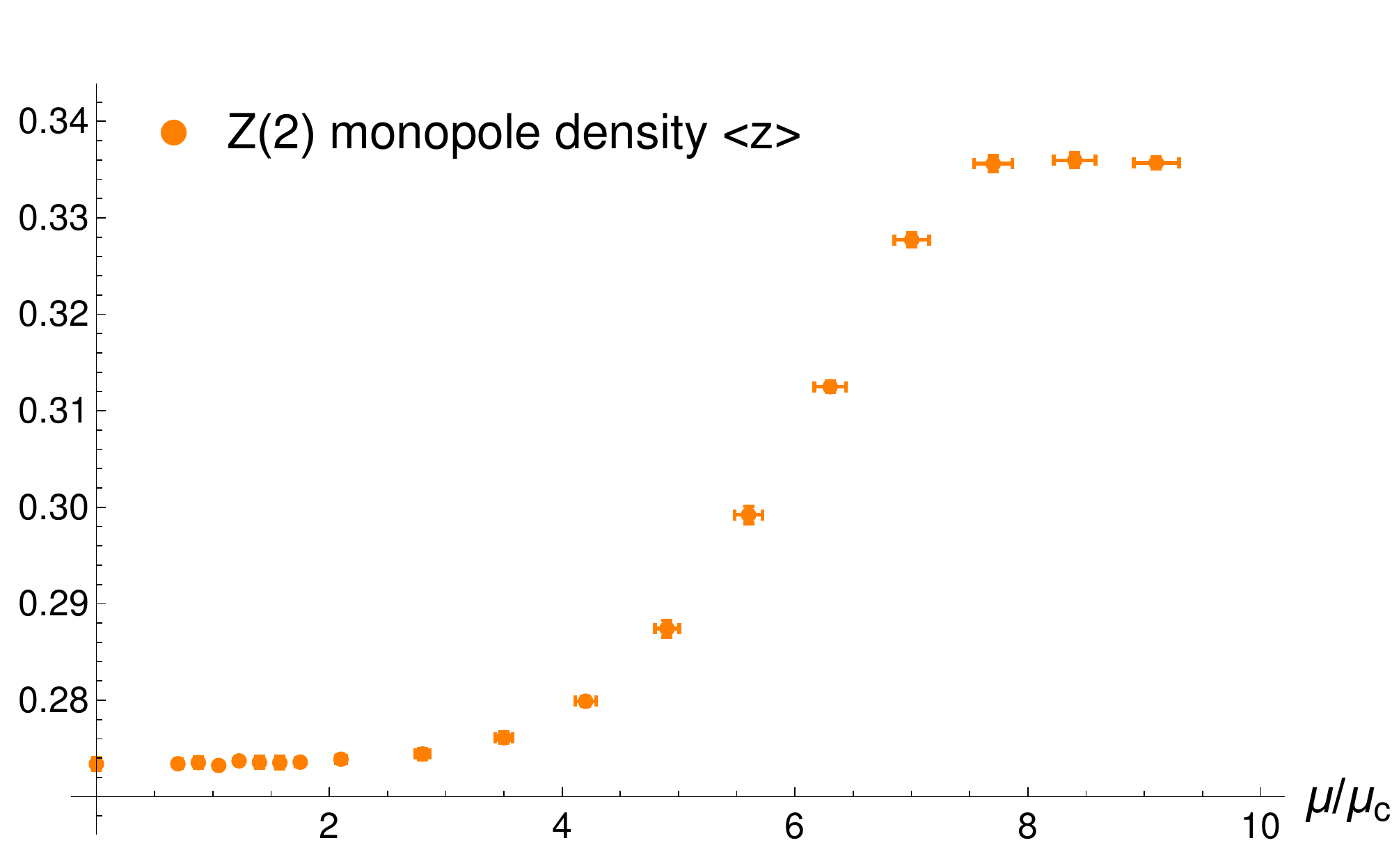}
	\caption{$Z_2$ monopole density for staggered quarks with $|\Lambda|=16^3\times32$, $a m=0.01$, $a \lambda = 0.001$ and $\beta=1.7$.}
	\label{fig:z2mono_stag}
\end{figure}
\begin{figure}
	\centering
	\includegraphics[width=0.5\textwidth]{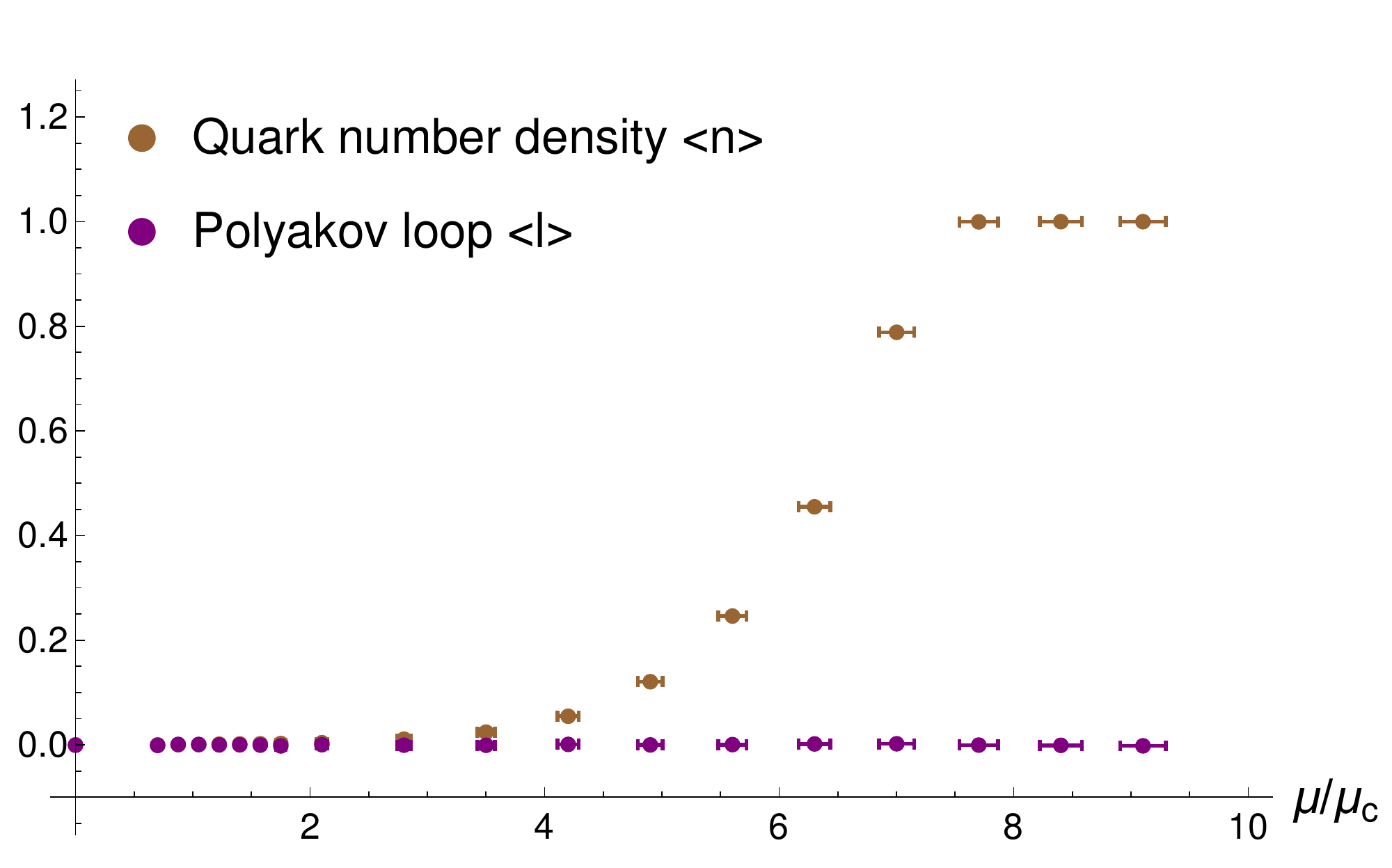}
	\caption{Quark number density and Polyakov loop for staggered quarks with  $|\Lambda|=16^3\times32$, $a m=0.01$, $a \lambda = 0.001$ and $\beta=1.7$.}
	\label{fig:ploop_qdens_stag}
\end{figure}

Measuring the $\mu$ dependence of the $Z_2$ monopole density and the quark number density, we observe that both
quantities saturate at large $\mu$ (see Figure~\ref{fig:z2mono_stag} and \ref{fig:ploop_qdens_stag}). In these figures, $\mu$ has been 
normalized with the critical chemical potential  $\mu_\text{c}=m_\pi/2$. With increasing $\mu$ the $Z_2$ monopole density 
approaches its quenched value, while at the same point the quark number density saturates. We conclude that in this high chemical 
potential regime the lattice is fully occupied with fermions, such that the system effectively becomes quenched.

Finally, we observe that the Polyakov loop is rather insensitive to the chemical potential 
with staggered quarks, and in fact coincides with its value in the quenched limit (see Figure \ref{fig:ploop_qdens_stag}).
This is in contrast to lattice simulations of two-color QCD with Wilson fermion \cite{Cotter:2012mb,Holicki:2017psk} and $G_2$-QCD with Wilson fermions \cite{Maas:2012wr}, where the Polyakov loop shows a peak around half filling.
Also, in a previous effective Polyakov loop model study for QCD-like theories \cite{Scior:2015vra,Scior:2016fso} with heavy Wilson quarks it has been seen that the Polyakov loop expectation value has a peak at the inflection point of the quark number density.

To explain this discrepancy, we consider that in two-dimensional two-color QCD, where large temporal extends of the lattice are feasible, the peak vanishes in the limit of $N_t\to \infty$ \cite{Wellegehausen:2017gba}, suggesting that the non-vanishing Polyakov loop might be an effect of
the residual temperature due to the finite lattice volume. This is also in agreement with recent lattice simulations with staggered fermions at zero temperature but larger inverse gauge coupling, leading to a larger residual temperature \cite{Nikolaev2017}. Here the Polyakov loop also increases with increasing chemical potential.
For Wilson fermions, a larger lattice spacing might lead to the excitement of heavy doublers beyond some $a\mu$, such that the free energy becomes finite. As these heavy doublers are not present in the staggered formalism, this behaviour is not observed here at comparable lattice spacings.

\section{Chiral Symmetry Breaking Pattern and the Goldstone Spectrum}\label{sec:patsymbreak}

Having established a set of parameters ($\beta=1.7$, $16^3\times 32$, $am=0.01$) with which we 
expect to reproduce the continuum physics, we now turn to the primary focus of this work which
is to study the Goldstone spectrum. We begin by reviewing the symmetry-breaking channels
of the staggered action of two-color lattice QCD and discussing the associated Goldstone modes
and correlation functions, 
which will then be compared inside and outside of the bulk phase. These issues were previously
discussed in Ref. \cite{Hands:1999md}. We present a compact summary
here to keep this paper self-contained.

For this purpose, it is convenient to introduce a new basis for the fermion fields, given by
\begin{equation}
\bar{X}_{e}=\left(\begin{array}{cc}
\bar{\chi}_{e}  &-\chi^{T}_{e}\tau_{2}
\end{array}\right)\ ,\ \ 
X_{o}=\left(\begin{array}{c}
\chi_{o}        \\-\tau_{2}\bar{\chi}^{T}_{o} 
\end{array}\right) \ ,
\label{eobasis}
\end{equation}
which separates even and odd sites. The kinetic part of the staggered action (\ref{eq:gorkovaction}) then reads \cite{Hands:1999md}
\begin{widetext}
\begin{equation}
S_{kin}=\sum_{n\in\Lambda^{\prime},\nu} \dfrac{\eta_{\nu}(n)}{2} \left[ \bar{X}_{e}(n) 
\left(\begin{array}{cc}
e^{\mu \delta_{\nu,4}}  &0\\
0       &e^{-\mu \delta_{\nu,4}}
\end{array}\right) U_{\nu}(n)X_{o}(n+\hat{\nu}) - \bar{X}_{e}(n) 
\left(\begin{array}{cc}
e^{-\mu \delta_{\nu,4}} &0\\
0       &e^{\mu \delta_{\nu,4}}
\end{array}\right)
U_{\nu}(n-\hat{\nu})^{\dagger}X_{o}(n-\hat{\nu}) \right]\ ,
\end{equation}
\end{widetext}
where the sum runs over even sites only. 
It follows that in the limit $m=\lambda=\mu=0$ the fermion action is invariant under 
\begin{equation}
X_{o} \rightarrow V X_{o}\ \ ,\ \bar{X}_{e} \rightarrow \bar{X}_{e} V^{\dagger}\ \ ;\ V \in U(2) \ .
\end{equation}
The original $U(1)_{e} \times U(1)_{o}$ symmetry of the $N_f=1$ staggered action 
for two-color QCD is therefore enlarged to $U(2)$ in this limit \cite{Hands:1999md,Kogut:2001na,Kogut:2003ju}.

Applying the same basis transformation to the mass and the diquark source terms one obtains
\begin{align}
\bar{\chi}\chi &= \frac{1}{2}\left[ \bar{X}_{e} 
\sigma_1 \tau_{2}\bar{X}_{e}^{T}
+ X_{o}^{T} 
\sigma_1
\tau_{2}X_{o}
\right]\ ,\label{qbqeo}\\
\chi \chi &= \frac{1}{2}\left[ \bar{X}_{e} 
\sigma_3 \tau_{2}\bar{X}_{e}^{T}
+ X_{o}^{T} 
\sigma_3
\tau_{2}X_{o}
\right]\ , \label{qqeo}
\end{align}
where we understand Pauli matrices $\sigma_i$ to act in the basis (\ref{eobasis})
and $\tau_i$ to act on color indices. Hence, the condensates are indistinguishable at $\mu=0$ as
they are connected by $V=\frac{i}{\sqrt{2}}({\bf 1}+i\sigma_2) \in U(2)$ \cite{Hands:1999md}. 
The Goldstone modes are derived by applying infinitesimal $U(2)$ rotations
\begin{equation}
V_{\delta} = {\bf 1} + i\delta \lambda \ \ ,\ \lambda \in \{{\bf 1},\tau_{i}\}
\end{equation}
to $\bar{\chi}\chi$ and $\chi\chi$, where the coefficient of $O(\delta)$ is then identified as the 
Goldstone mode \cite{Hands:1999md}. The results are shown in Table \ref{tab:gm}.

Both condensates leave one generator of $U(2)$ unbroken and hence induce the same symmetry-breaking pattern $U(2)\rightarrow U(1)$ at $\mu=0$. 
Since at $\mu \neq 0$ the symmetry is reduced from $U(2)$ to $U(1)_{e} \times U(1)_{o}$, one is left with two generators 
$\left\{ {\bf 1}, \sigma_{3} \right\}$ of $U(2)$, which correspond to the staggered $U(1)_{\epsilon}$ and baryon number conservation, respectively.
\begin{table}[htb]
\centering
\renewcommand{\arraystretch}{1.5}
\begin{tabular}{c|c|c}
\hspace{0.3cm}	&$\left< \bar{q}q\right>\hspace{0.3cm}$ &$\left< q q\right>$\\%[0.2cm]
	\hline%\\[-0.3cm]
${\bf 1}$\hspace{0.3cm} &$\bar{\chi}\epsilon\chi$\hspace{0.3cm} &$\chi^{T}\tau_{2}\epsilon\chi + \bar{\chi}\tau_{2}\epsilon\bar{\chi}^{T}$\\%[0.2cm]
$\tau_{1}$\hspace{0.3cm} &$\chi^{T}\tau_{2}\chi - \bar{\chi}\tau_{2}\bar{\chi}^{T}$\hspace{0.3cm} &-\\%[0.2cm]
$\tau_{2}$\hspace{0.3cm} &$\chi^{T}\tau_{2}\chi + \bar{\chi}\tau_{2}\bar{\chi}^{T}$\hspace{0.3cm} &$\bar{\chi}\chi$\\%[0.2cm]
$\tau_{3}$\hspace{0.3cm} &-\hspace{0.3cm} &$\chi^{T}\tau_{2}\chi - \bar{\chi}\tau_{2}\bar{\chi}^{T}$
\end{tabular}
\renewcommand{\arraystretch}{1}
\caption{The Goldstone modes for $N_\text{f}=1$ according to the generators of $U(2)$, where $\epsilon$ corresponds to \text{$\epsilon(n)=\eta_{5}(n)=(-1)^{n_{1}+n_{2}+n_{3}+n_{4}}$}.}
\label{tab:gm}
\end{table}

\begin{figure}[htb]
\begin{tikzpicture}[align=center,->,>=stealth']
  \SetGraphUnit{5}
  \node[](1) {$U(2N_{f})$};
  \node[node distance=5cm](2)[right of=1] {$O(2N_{f})$};
  \node[node distance=2cm](3)[below of=1] {$U(N_{f})\times U(N_{f})$};
  \node[node distance=2cm](4)[below of=2] {$U(N_{f})\times U(1)_{B}$};
  \node[node distance=2cm](5)[below of=4] {$O(N_{f})$};
  \path[every node/.style={anchor=south}]
   (1) edge node {$\sqrt{\left<\bar{q}q\right>^{2}+\left<q q\right>^{2}}$} (2)
   (1) edge node[right] {$\mu\neq 0$} (3)
   (4) edge node[right] {$\left<q q\right>$} (5)
   (3) edge node[above right] {$\left<q q\right>$} (5)
   (3) edge node {$\left<\bar{q}q\right>$} (4);
\end{tikzpicture}
\caption{Symmetry-breaking pattern of staggered action for two-color QCD at $N_f\geq 1$.  
\label{fig:multiflavor_sb}}
\end{figure}
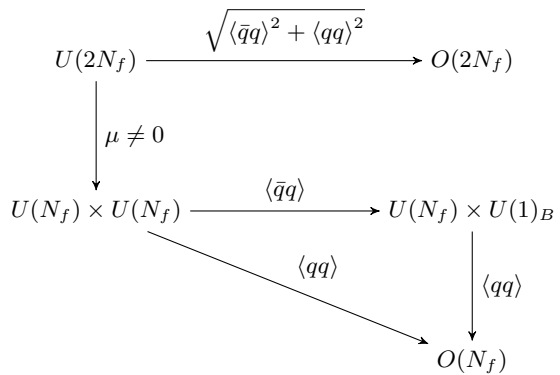

The above can be generalized to $N_{f}>1$ staggered fermions. There one has a $U(1)\times U(1)$ 
symmetry for each flavor, leading to $U(N_f)\times U(N_f)$, which is extended to $U(2N_f)$ at
$\mu=0$. The same Goldstone modes listed in Table \ref{tab:gm} appear also for $N_{f}>1$ (and
in particular for the $N_{f}=2$ case considered in this paper), but with different multiplicities.
 The full symmetry-breaking pattern is summarized in Fig. \ref{fig:multiflavor_sb}.
It can be seen that any-color QCD with quarks in the adjoint representation in the continuum exhibits the same pattern of
symmetry breaking, with an additional breaking of $U(1)_{A}$ due to the axial anomaly \cite{Kogut:2000ek}.

The Goldstone spectrum consists of two meson modes, the (pseudoscalar) pion $\pi$ and the scalar meson $f_{0}$, and two diquark modes, a scalar diquark $qq$ 
and a pseudoscalar diquark $\epsilon qq$. 
In Table \ref{fig:interpol1}, we show the employed interpolating operators for the $f_{0}$ and $\pi$ modes taken from Ref. \cite{ALTMEYER1993445}.
It is important to realize that channel 1 not only contains the desired scalar meson, but also an excited pion. However, these two states can be 
separated during the fitting procedure as they have opposite parity. The groundstate pion is exclusively contained in channel 2.
The interpolating operators from \cite{Kogut:2003ju} are employed for the (pseudo-)scalar diquark modes, shown in Table \ref{fig:interpol2}
These modes furthermore contain contributions from their corresponding anti-diquarks. Nevertheless, the anti-diquark modes become less important with 
increasing chemical potential as the propagation of particles is favored ($e^{a\mu}$) over the propagation of anti-particles ($e^{-a\mu}$) at non-vanishing 
chemical potential.

\begin{table}[htb]
\centering
\renewcommand{\arraystretch}{1.5}
\begin{tabular}{cccc}
Channel &Operator       &$J^{PC}$       &States\\
\hline
1       &$\bar{\chi}\chi$       &$0^{++}$       &$f_{0}$\\
        &       &$0^{-+}$       &$\pi$\\
\hline
2       &$\eta_{4}\bar{\chi}\chi$       &$0^{+-}$       &-\\
        &       &$0^{-+}$       &$\pi$\\
\hline
\end{tabular}
\renewcommand{\arraystretch}{1}
\caption{The interpolating operators for $f_{0}$ and $\pi$ modes.\label{fig:interpol1}}
\end{table}

\begin{table}[htb]
\centering
\renewcommand{\arraystretch}{1.5}
\begin{tabular}{cccc}
Channel &Operator       &States\\
\hline \vspace{0.3cm}
3       &$\frac{1}{2}\left(\chi^{T}\tau_{2}\chi-\bar{\chi}\tau_{2}\bar{\chi}^{T}\right)$        &$qq$/$\bar{q}\bar{q}$\\
\hline \vspace{0.3cm}
4       &$\eta_{5}\frac{1}{2}\left(\chi^{T}\tau_{2}\chi+\bar{\chi}\tau_{2}\bar{\chi}^{T}\right)$        &$\varepsilon qq$/$\varepsilon\bar{q}\bar{q}$\\
\hline
\end{tabular}
\renewcommand{\arraystretch}{1}
\caption{The interpolating operators for the (pseudo-) scalar diquark modes. \label{fig:interpol2}}
\end{table}

To extract the ground state masses of the particle states, we employ the zero-momentum projected correlations functions of the form
\begin{equation}
C(t) = \sum_{\vec{x}} \left\langle 0\left|O(\vec{x},t) \bar{O}(\vec{0},0)\right|0\right\rangle \ .
\end{equation}
Note that only the connected contributions are considered here. For the different channels shown
in Tables \ref{fig:interpol1} and \ref{fig:interpol2} we obtain:
\begin{itemize}
\item Channel 1 - Scalar Meson
\begin{equation}
C(t) = -\sum_{\vec{x}} \eta_{5}(\vec{x},t)\ \text{tr}\left[ G^{\dagger}[-\mu](\vec{x},t;0)G[\mu](\vec{x},t;0) \right]
\end{equation}
\item Channel 2 - Pion / Pseudoscalar Meson
\begin{equation}
C(t) = -(-1)^{t}\sum_{\vec{x}} \text{tr}\left[ G^{\dagger}[-\mu](\vec{x},t;0)G[\mu](\vec{x},t;0) \right]
\end{equation}
\item Channel 3 - Scalar Diquark
\begin{align}
\begin{split}
C(t) = & \frac{1}{2} \sum_{\vec{x}} \left\{ \text{tr}\left[ G^{T}[\mu](\vec{x},t;0)\tau_{2}G[\mu](\vec{x},t;0)\tau_{2} \right] \right.\\ &\left. + \text{tr}\left[ G^{\dagger}[-\mu](\vec{x},t;0)\tau_{2}(G^{\dagger})^{T}[-\mu](\vec{x},t;0)\tau_{2} \right] \right\} \label{channel3}
\end{split}
\end{align}
\item Channel 4 - Pseudoscalar Diquark
\begin{align}
\begin{split}
C(t) =& \frac{1}{2} \sum_{\vec{x}} \eta_{5}(\vec{x},t) \left\{ \text{tr}\left[ G^{T}[\mu](\vec{x},t;0)\tau_{2}G[\mu](\vec{x},t;0)\tau_{2} \right] \right.\\ &\left. + \text{tr}\left[ G^{\dagger}[-\mu](\vec{x},t;0)\tau_{2}(G^{\dagger})^{T}[-\mu](\vec{x},t;0)\tau_{2} \right] \right\}
\end{split}
\end{align}
\end{itemize}
We used the notation $G=\left(D^{\dagger}D+\lambda^{2}\right)^{-1}D^{\dagger}$ here, 
which corresponds to only the off-diagonal terms in Eq. (\ref{eq:full_prop}). The 
diagonal terms produce corrections of order $O(\lambda^{2})$, which we refrain from
listing here explicitly as their contributions are negligibly small for the values
of $\lambda$ considered in this work. We did in fact include these corrections for the
results shown in Figs. \ref{fig:gsmb1.5} and \ref{fig:gsmb1.7}. The results in Fig. 
\ref{fig:gsmb1.7pion} were obtained without them. 

\subsection{Goldstone spectrum in the bulk phase}\label{sec:spectrum_bulk}

We now study the (pseudo) Goldstone spectrum on a $12^{3}\times 24$ lattice at $\beta=1.5$ with 
quark mass $a m=0.025$ and diquark source $\lambda=0.0025$.
As the combined condensate $\sqrt{\left<\bar{q}q\right>^{2}+\left<q q\right>^{2}}$ rotates from a 
chiral to a diquark condensate with increasing chemical potential,
we assume that the Goldstone modes corresponding to a given $U(2)$ generator mix in a similar way, i.e. rotate into each other 
with the same rotation angle $\alpha(\mu)$, as described by Eq.~(\ref{eq:xpt_angle}).
Hence, we introduce the two combined modes
\begin{itemize}
\item[ ] $\bar{q}\bar{q}/f_{0}$:\ \ $\frac{1}{2}\left(\chi^{T}\tau_{2}\chi + \bar{\chi}\tau_{2}\bar{\chi}^{T}\right) \cos\alpha + \bar{\chi}\chi \sin\alpha$
\item[ ] $\pi/\epsilon qq$:\ \ $\bar{\chi}\epsilon\chi \cos\alpha + \frac{1}{2}\left(\chi^{T}\tau_{2}\epsilon\chi + \bar{\chi}\tau_{2}\epsilon\bar{\chi}^{T}\right) \sin\alpha$ .
\end{itemize}
Note that this simple mode mixing holds only to leading order $\chi$PT.
Note also that in addition to the terms of order $O(\lambda^{2})$
discussed above, the correlators of the combined modes $\bar{q}\bar{q}/f_0$ and $\pi/\epsilon qq$ 
contain additional terms of order $O(\lambda)$ from the mixed parts (i.e. the terms $\sim\sin\alpha\cos\alpha)$,  
as diagonal elements of the full propagator in Nambu-Gorkov space \eqref{eq:full_prop} contribute 
to these. We again refrain from listing these terms here explicitly but included them in our simulations. 

To obtain the masses of the (pseudo) Goldstone modes, we measure the zero-momentum projected connected correlation 
functions and extract their masses from fits to $\cosh(\cdot)$ (which combines exponential decays forwards and backwards 
in Euclidean time). For $\pi/\epsilon qq$ a factor $(-1)^t$ is inserted to account for negative parity.
In the case of the $\bar{q}\bar{q}/f_{0}$ mode we have to use the fitting function
\begin{align}
\begin{split}
C(t) =\ &A\ \cosh\left( m_{qq}\left(t-\frac{N_{t}}{2}\right) \right) \\ &+  B\ \cosh\left( m_{\bar{q}\bar{q}/f_{0}}\left(t-\frac{N_{t}}{2}\right) \right)
\end{split}
\end{align}
as the correlator contains a contribution from the scalar diquark mode $qq$ in addition to the mixing of the scalar anti-diquark $\bar{q}\bar{q}$ with the scalar meson $f_{0}$.

At $\mu \neq 0$ the combined mode $\bar{q}\bar{q}/f_{0}$ is a massive pseudo Goldstone mode for all values of $\mu$.
Including a non-vanishing quark mass $m \neq 0$, also the combined mode $\pi/\epsilon qq$ becomes a pseudo Goldstone mode as the $U(1)_{A}$ symmetry gets broken.
The only true Goldstone mode is given by the scalar diquark mode $qq$ in the limit $\lambda\rightarrow 0$ and for $\mu>\mu_\text{c}$.
Generally for $\mu<\mu_\text{c}$ and $\lambda \rightarrow 0$, the pion mass $m_\pi$ stays constant as the pion does not carry a net Baryon number,
whereas the scalar diquark mass $m_{qq}$ decreases like $m_{\pi}-2\mu$ and the scalar anti-diquark mass 
$m_{\bar{q}\bar{q}}$ increases like $m_{\pi}+2\mu$. 

\begin{figure}[htb]
\includegraphics[width=0.5\textwidth]{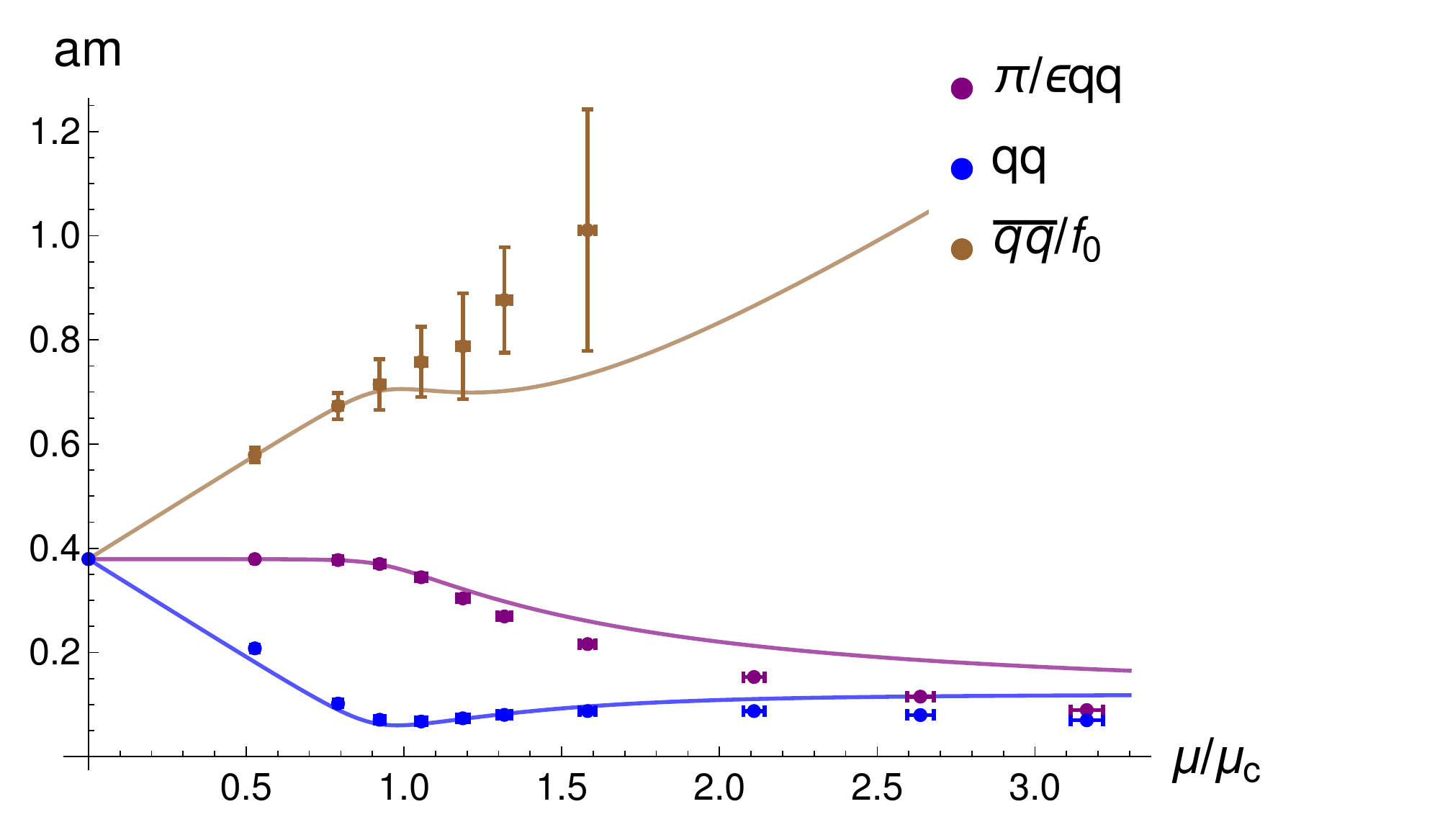}
\caption{The (pseudo) Goldstone spectrum on a $12^{3}\times 24$ lattice at $\beta=1.5$ with quark mass $a m=0.025$ and diquark source $\lambda=0.0025$.}
\label{fig:gsmb1.5}
\end{figure}

We compare our obtained masses of the (pseudo) Goldstone modes to the corresponding $\chi$PT predictions for twocolor
QCD with staggered quarks \cite{Kogut:2003ju}, using $\mu_c$ from the fit of the condensates in Section~\ref{sec:xpt} (see Figure~\ref{fig:gsmb1.5}) 
and the lattice parameters $a\lambda$ and $a m$ as input.
The large error of the combined mode $\bar{q}\bar{q}/f_0$ mainly comes from the systematic error as the double-cosh fit is more sensitive to the fitting interval than the single-cosh fits of the other modes.

We find a general agreement of the scalar diquark mode $qq$ and the $\pi/\epsilon qq$ mode to their predictions.
Deviations become notable at large chemical potential, where also the quark number density in Figure~\ref{fig:xpt_fit} deviates largely.
The large discrepancy of the $\bar{q}\bar{q}/f_{0}$ mode for $\mu\geqslant\mu_\text{c}$ might be due to omitting  disconnected contributions,
as at large chemical potential the scalar meson mode $f_{0}$ dominates this combined mode.
Note that the scalar diquark mode $qq$ has disconnected contributions of order $O(\lambda^2)$, which we expect to have a small effect.
Hence, we obtain similar results as in the previous study \cite{Kogut:2003ju} for $N_\text{f}=4$.
However, in Ref. \cite{Kogut:2003ju} disconnected contributions have not been omitted and thus their obtained $\bar{q}\bar{q}/f_{0}$ mode coincides better with the $\chi$PT prediction.
In conclusion, we find that within the bulk phase fundamental staggered quarks resemble the chiral symmetry breaking pattern of adjoint QCD or $G_2$-QCD in the continuum as most notably seen
in the behavior of the pion branch above the onset at $a \mu_\text{c}$.

\subsection{Goldstone spectrum outside bulk phase}\label{sec:spectrum_nonbulk}

Continuum two-color QCD with quarks in the fundamental representation obeys the pattern of symmetry breaking $SU(2N_\text{f})\rightarrow Sp(2N_\text{f})$.
Applying $\chi$PT, it was found that the mass of the pion mode increases for $\mu >\mu_\text{c}$ \cite{Kogut:2000ek} in this
case, due to a swapping of the $P_S$ and $P_A$ branches when compared to the staggered 
action \cite{Kogut:2003ju}. 
We now wish to test whether the pattern of symmetry breaking on the lattice will change to the 
continuum pattern in the limit $a\rightarrow 0$. Thus, we calculate the spectrum
again at a larger inverse gauge coupling and using an improved gauge action.

With the new parameters ($\beta=1.7$, $N_s=16$ and $N_t=32$), where monopoles are strongly suppressed, we find a quite 
different behaviour of the (pseudo) Goldstone modes than in Sec. \ref{sec:spectrum_bulk}.
We observe that the combined modes $\bar{q}\bar{q}/f_{0}$ and $\pi/\epsilon qq$ do not give any meaningful results.
Instead, we therefore study the scalar diquark mode $qq$ and the pion mode $\pi$ individually, and additionally we extract 
the mass of the scalar anti-diquark $\bar{q}\bar{q}$ from the operator 
$\chi^{T}\tau_{2}\chi-\bar{\chi}\tau_{2}\bar{\chi}^{T}$ for $\mu<\mu_\text{c}$.
In the correlation function of the pion mode we find an additional contribution from an opposite parity state, which we 
filter out by applying a double cosh-fit of the form
\begin{align}
\begin{split}
C(t) =\ &A\ \cosh\left( m_{\pi}\left(t-\frac{N_{t}}{2}\right) \right) \\ &+  (-1)^{t}B\ \cosh\left( m_{\pi^{\star}}\left(t-\frac{N_{t}}{2}\right) \right)\ .
\end{split}
\end{align}

\begin{figure}[h]
\includegraphics[width=0.5\textwidth]{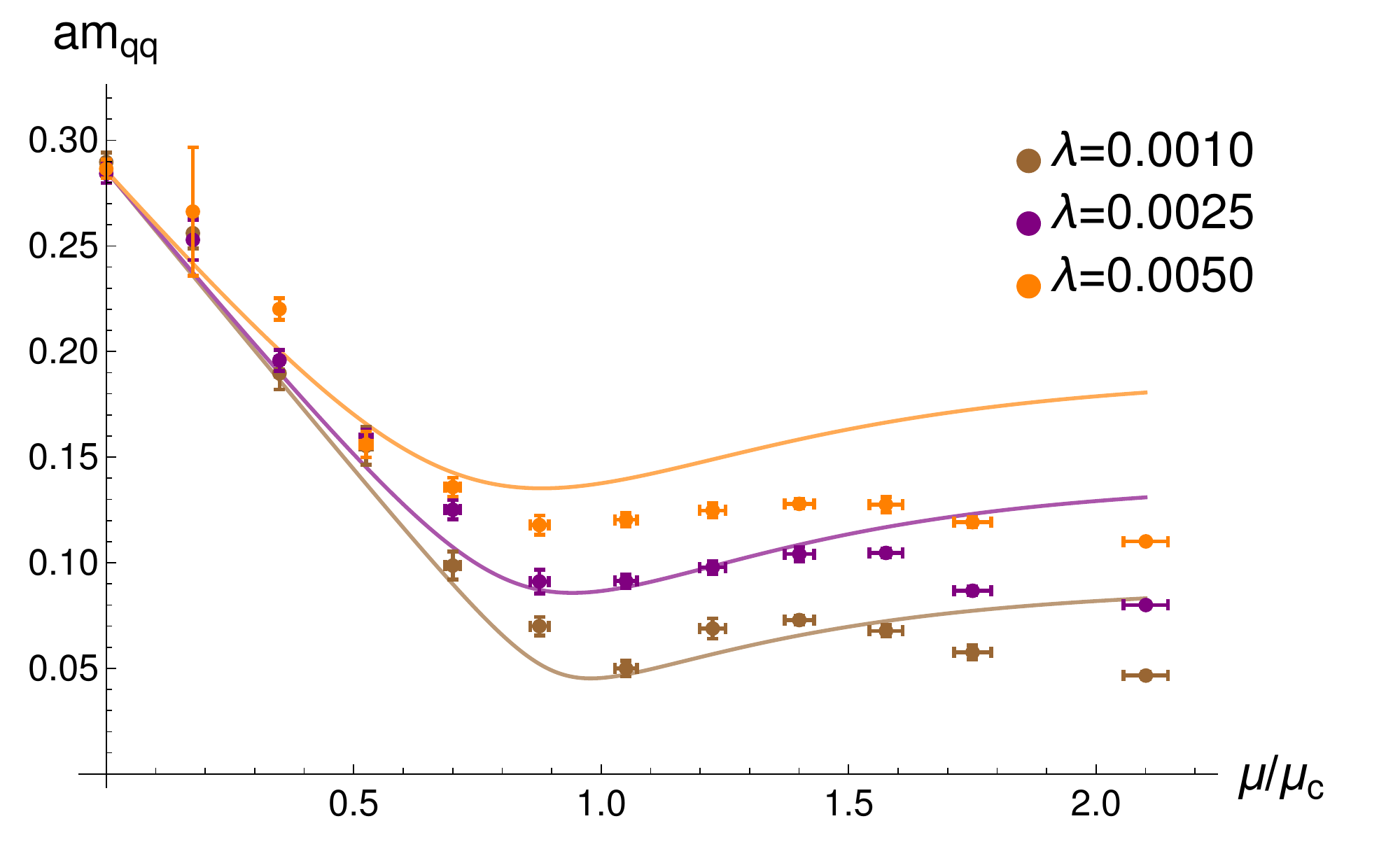}
\includegraphics[width=0.5\textwidth]{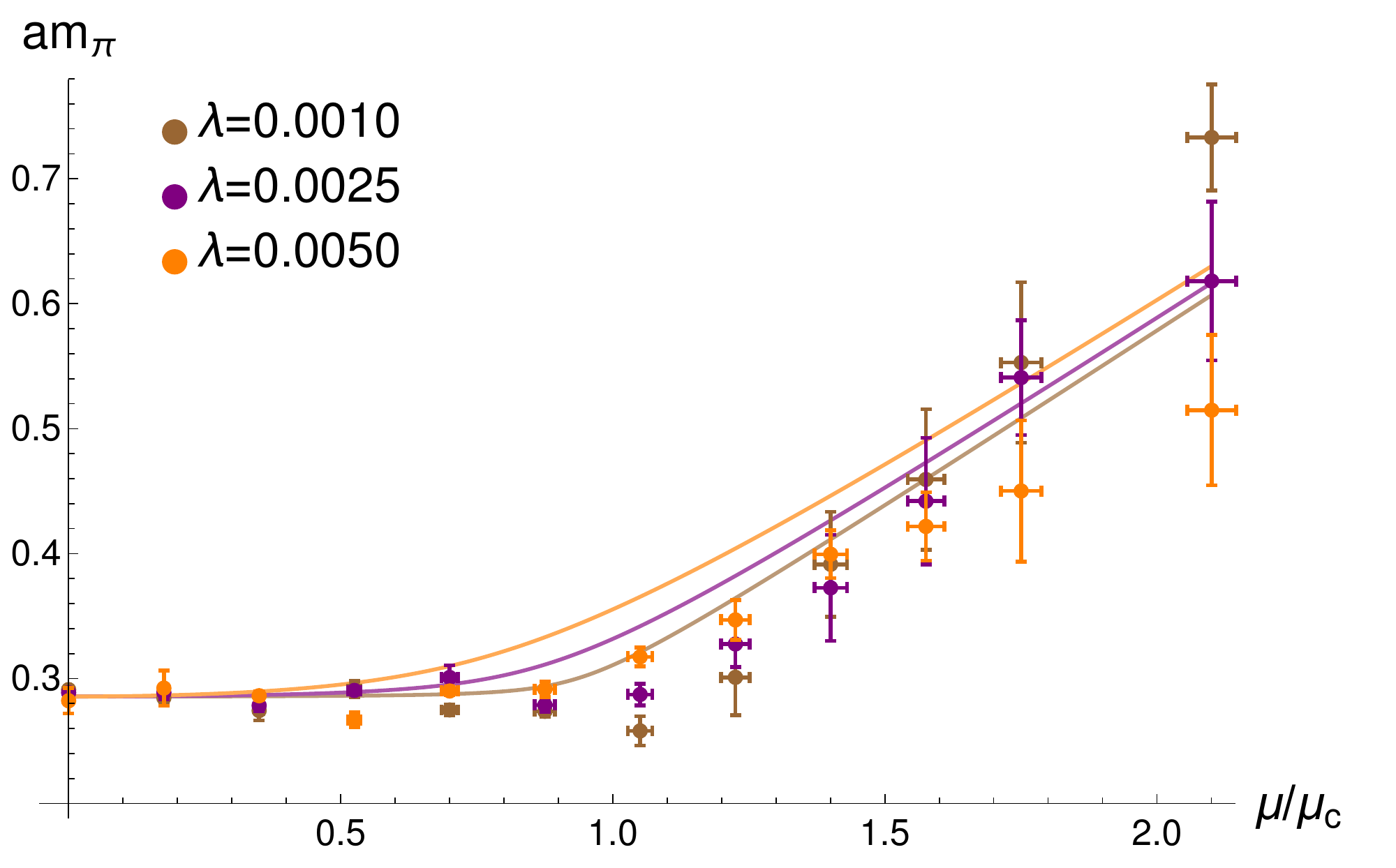}
\caption{The scalar diquark mass (top) and the pion mass (bottom) on a $16^{3}\times 32$ lattice at $\beta=1.7$ with quark mass $a m=0.01$ for different diquark sources $\lambda$.}
\label{fig:gsmb1.7pion}
\end{figure}

We first study the $\lambda$-dependence pion mass and the scalar diquark mass (see Figure~\ref{fig:gsmb1.7pion}). Here we neglected 
$O(\lambda^{2})$ contributions from the diagonal terms in Eq. (\ref{eq:full_prop}) in
the calculation of the correlation functions.
We compare the results to $\chi$PT predictions for from continuum two-color QCD with 
quarks in the fundamental representation, for which we use the lattice 
parameters $\lambda$ and $m$, and also $\mu_\text{c}$ as obtained from measuring the pion mass at $\mu=0$, as input. 
For the smallest diquark source $\lambda=0.001$ we find excellent agreement with $\chi$PT up to $\mu\sim1.7\mu_c$
where leading order $\chi$PT is not reliable anymore. We conclude that the diagonal terms are negligible for this
choice of diquark source. 

\begin{figure}[h]
\includegraphics[width=0.5\textwidth]{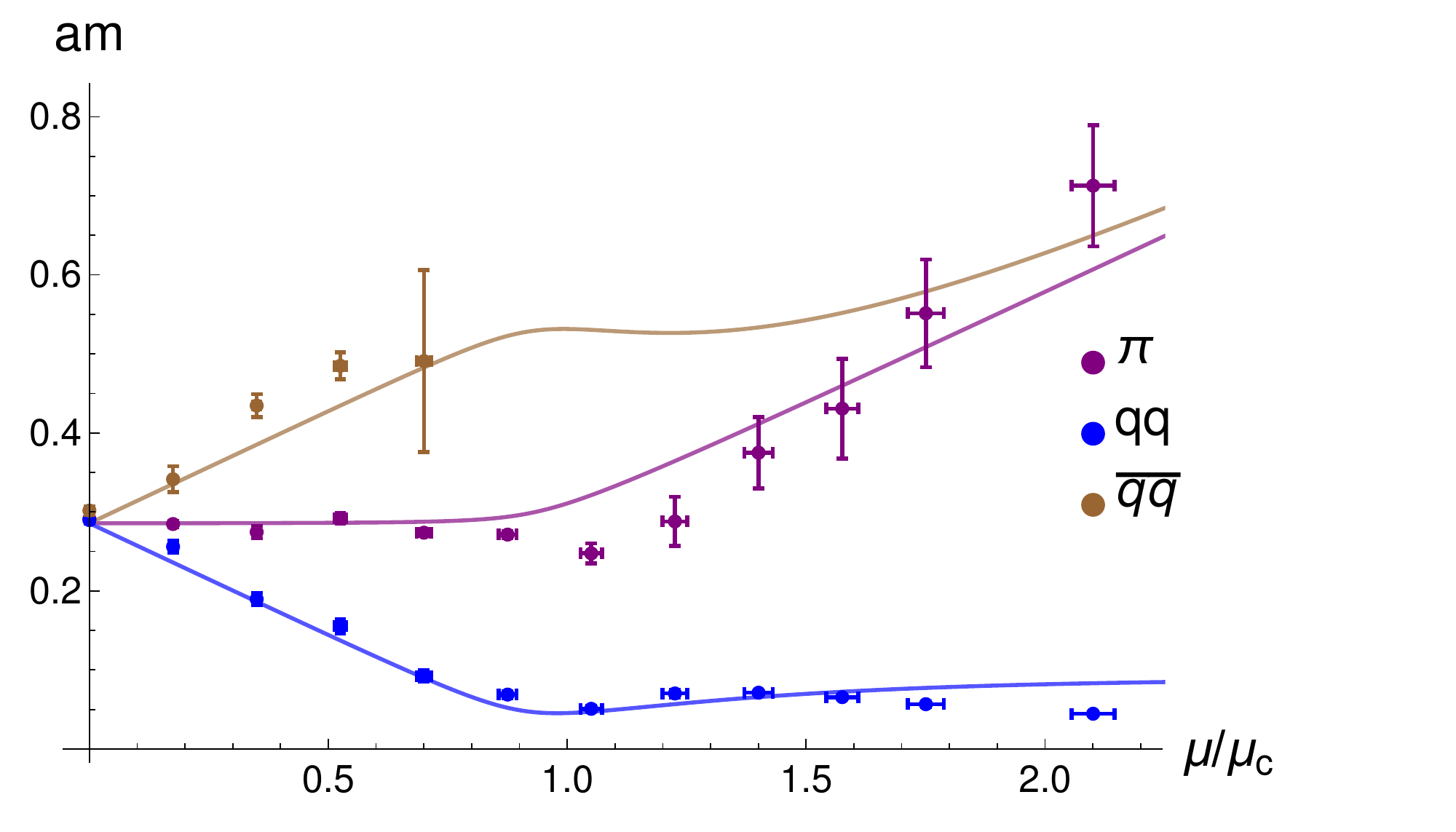}
\caption{The (pseudo) Goldstone spectrum on a $16^{3}\times 32$ lattice at $\beta=1.7$ with quark mass $a m=0.01$ and diquark source $\lambda=0.001$.}
\label{fig:gsmb1.7}
\end{figure}

At $\beta=1.7$ the pion mass stays constant for $\mu\leqslant\mu_\text{c}$, but it starts to increase for $\mu>\mu_\text{c}$. 
This differs  profoundly from the behaviour at $\beta=1.5$ (cf. Figure \ref{fig:gsmb1.5}). 
Again, the large error of the pion mode $\pi$ for $\mu>\mu_\text{c}$ 
mainly comes from the sensitivity of the double-cosh fit to the fitting interval.
We interpret the increasing pion mass for $\mu>\mu_\text{c}$ as strong evidence that the pattern of symmetry breaking 
changed to its continuum counterpart outside of the bulk phase. 
Figure~\ref{fig:gsmb1.7} combines the results for all the considered meson channels, for $\lambda=0.001$ where the agreement 
with $\chi$PT is nearly perfect. In this figure the $O(\lambda^{2})$ diagonal terms were in fact included, but their contribution is of similar magnitude
as the statistical error. 

\section{Unfolded level spacings}\label{sec:levelspac}

\begin{figure}[h]
\includegraphics[width=0.5\textwidth]{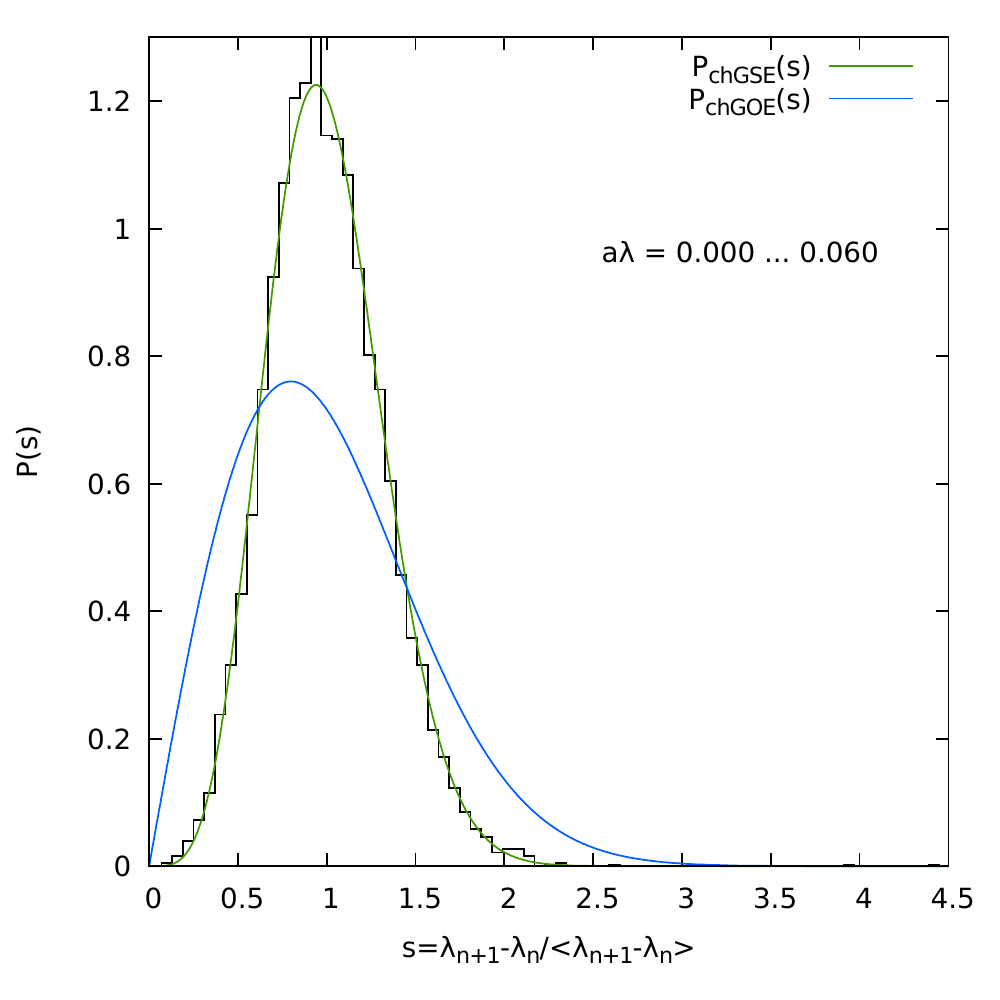}
\caption{The unfolded level spacing distribution of the staggered Dirac operator spectrum 
inside the bulk phase (unimproved action, $12^3\times24$, $\beta=1.5$). The level spacings are entirely 
distributed according to the RMT prediction for the chiral symplectic ensemble.}
\label{fig:ulsWilson1.5}
\end{figure}

\begin{figure}[h]
\includegraphics[width=0.5\textwidth]{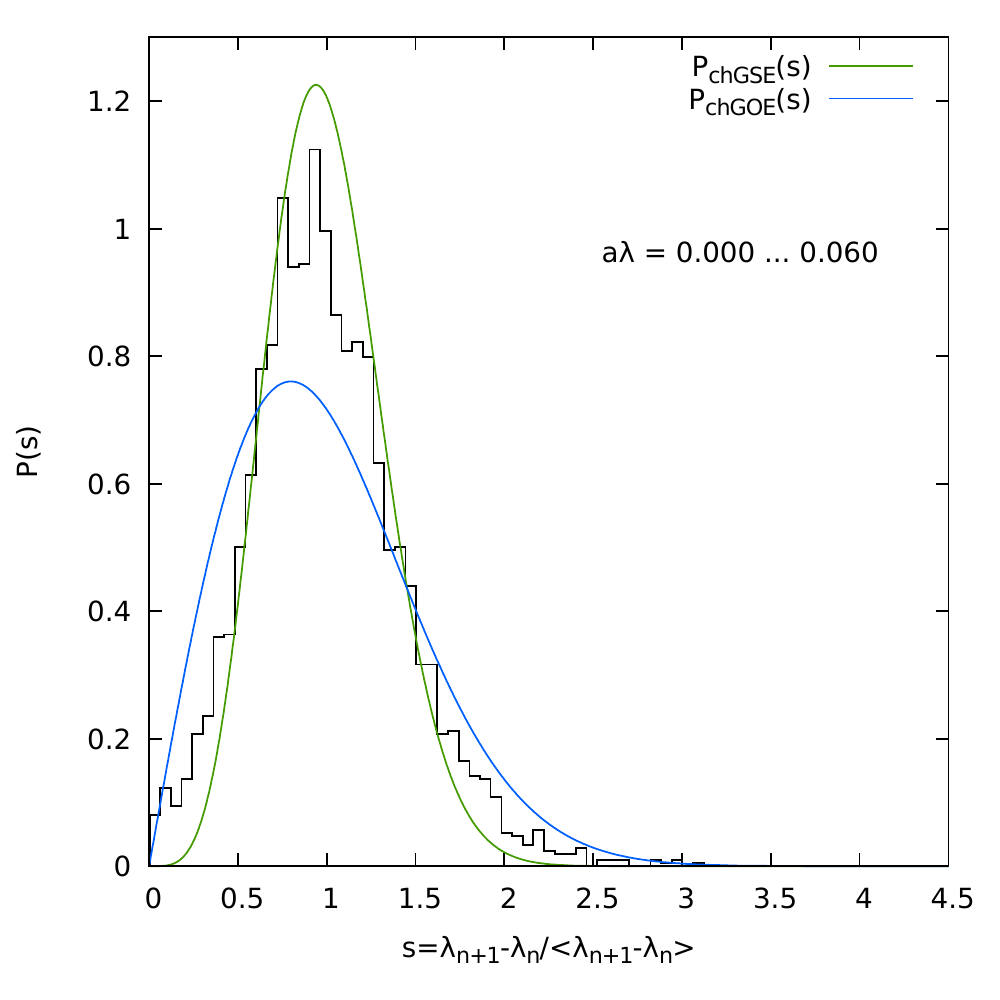}
\caption{The unfolded level spacing distribution of the staggered Dirac operator 
outside the bulk phase (Symanzik action, 
$16^3\times32$, $\beta=1.7$). Neither GSE nor GOE describe the distribution entirely.}
\label{fig:ulsSymanzik1.7}
\end{figure}

\begin{figure}[h]
\includegraphics[width=0.5\textwidth]{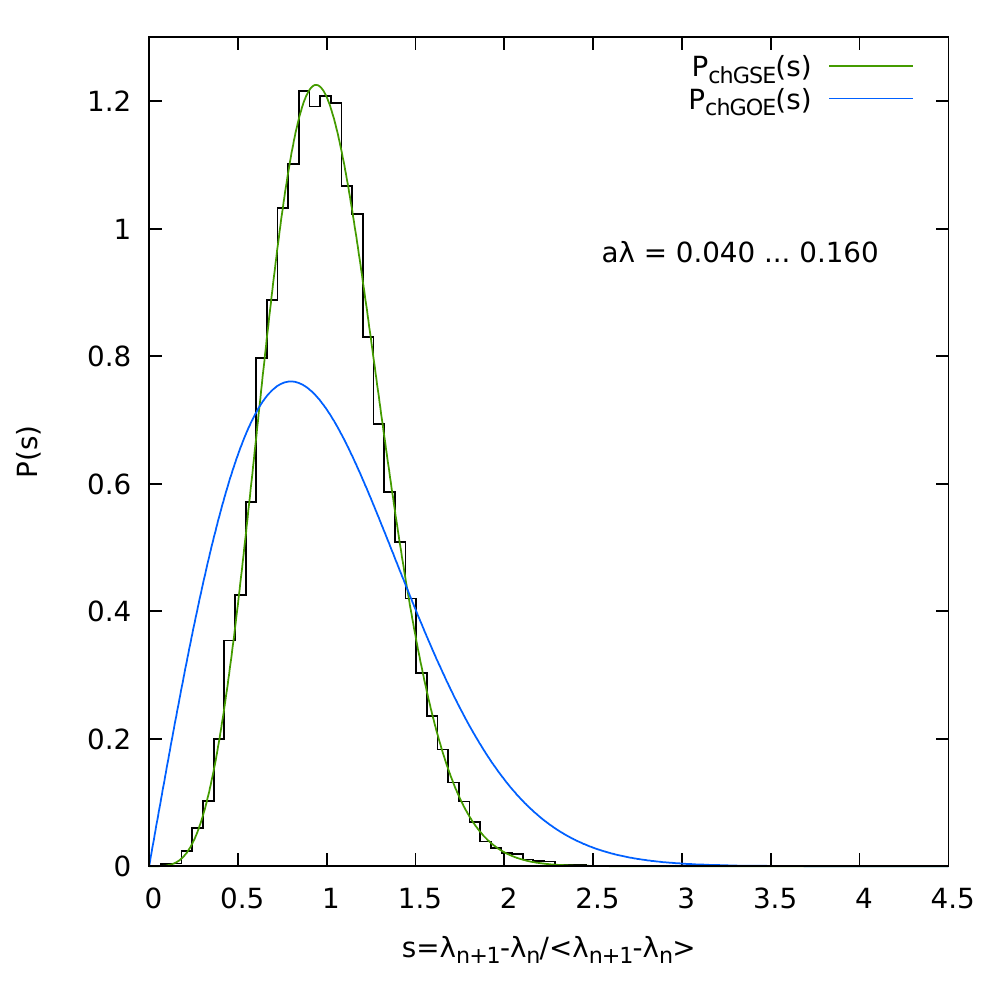}\\
\includegraphics[width=0.5\textwidth]{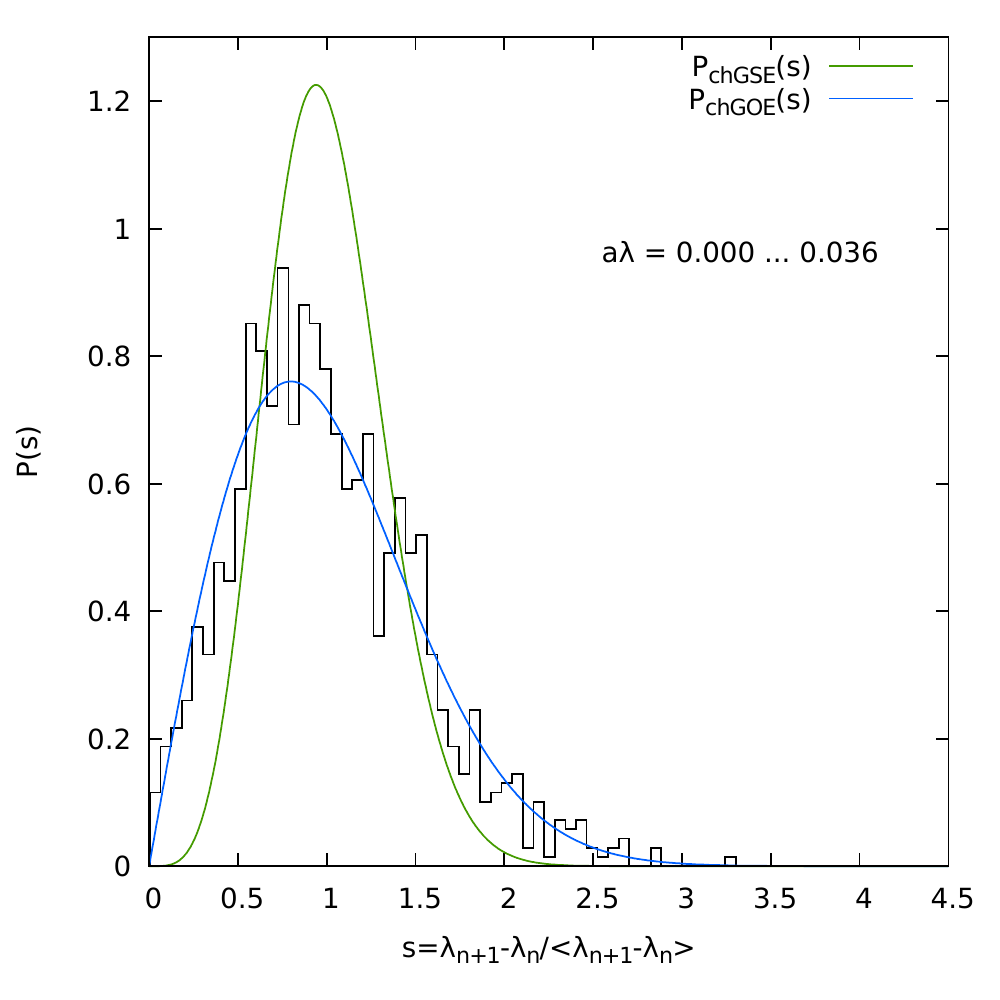}
\caption{The unfolded level spacing distribution of the staggered Dirac operator 
outside the bulk phase (Symanzik action, 
$16^3\times32$, $\beta=1.7$). Low and high modes shown separately. 
High-lying modes (top) are described by GSE while low-lying modes (bottom) are described by GOE.}
\label{fig:ulsSymanzik1.7lowVShigh}
\end{figure}

The low energy features of QCD are dominated by its global symmetries. In this limit, QCD is well described with only the two lightest quarks. 
Low-energy QCD can be approximated with a random matrix theory very well, where all interactions among the degrees of freedoms of the 
theory are equally likely and completely determined by the global symmetries. In a random matrix theory (RMT), the matrix elements of the 
Dirac operator are replaced with uncorrelated random numbers, in a way such that the global symmetries of the operator prevail 
\cite{Verbaarschot:2000dy}. Any observable is then an average over random matrix elements and depends only on universal features of the theory, 
and not on microscopic interactions.

RMT has found wide application for studying the eigenvalue statistics in nuclear resonances \cite{metha2004}. The distribution of level 
spacings is cleared off of any microscopic interactions by rescaling the spacings such that their average is unity and removing the fluctuating 
part of the cumulative spectral function. This procedure is called \textit{unfolding}. We wish to demonstrate here that the change of the Goldstone 
spectrum of twocolor QCD with staggered fermions when leaving the bulk phase is accompanied by a change of the unfolded level spacing distribution 
(ULSD) of the Dirac operator.

For different random matrix ensembles different forms of ULSD have been predicted \cite{Bruckmann:2008xr, ChauHuuTai:2001xp}. For the Gaussian 
orthogonal ensemble (GOE), where the matrix elements are real and the Dyson index is $\beta_D=1$, the ULSD is
\begin{equation}
	P_{\beta_D=1}(s)=\frac \pi 2 s e^{ -\frac \pi 4 s^2 } .
\label{eq:goe}
\end{equation}
For the Gaussian symplectic ensemble (GSE) the matrix elements are quaternion real and the Dyson index is 
$\beta_D=4$. The ULSD of the GSE is given by
\begin{equation}
	P_{\beta_D=4}(s)=\left(\frac{8}{3}\right)^6 \frac{1}{\pi^3} s^4 e^{ -\left(\frac 8 3\right)^2 \frac 1 \pi s^2 } .
\label{eq:gse}
\end{equation}
Using the \emph{Implicitly Restarted Arnoldi Algorithm}, we have measured 
the ULSD both with the parameters from Ref. \cite{Kogut:2003ju} ($12^3\times24$, $am=0.025$, $\beta=1.5$, Wilson action)
within the bulk phase, as well as with our improved parameters ($16^3\times 32$, $am=0.01$ $\beta=1.7$, Symanzik action)
on the weak-coupling side of the bulk crossover transition.

The ULSD within the bulk phase are shown in Figure \ref{fig:ulsWilson1.5}.
We find that the entire spectral range is well described by Eq. (\ref{eq:gse}),
i.e. the distributions resembles the symplectic ensemble very well.
Inside the bulk phase, the ULSD does not seem to contain any component distributed according to the
orthogonal ensemble. 

In the continuum limit the spectrum is expected to resemble the Gaussian orthogonal ensemble. 
With our improved parameters, we find however that neither Eq. (\ref{eq:gse}) nor Eq. (\ref{eq:goe}), 
fully describe the ULSD. In fact, our numerical data (seen in Figure \ref{fig:ulsSymanzik1.7})
indicate some sort of intermediate state. When separating the low-lying ($a\lambda=0.000\ldots 0.036$)
from the high-lying eigenmodes ($a\lambda=0.040\ldots 0.160$) it becomes clear that 
a large part of the low eigenmodes are now distributed according to the GOE, while the higher eigenmodes 
remain distributed according to the GSE (see Figure \ref{fig:ulsSymanzik1.7lowVShigh}).
Since the low eigenmodes govern the Goldstone modes, it is clear 
that our spectroscopic results should reflect the  chiral-symmetry breaking pattern of the GOE.
We suspect that in the continuum limit the symplectic part will vanish entirely.

\section{Conclusion and Outlook}\label{sec:concandoutl}
In this work we investigated the influence of bulk effects in QC$_2$D with 
staggered fermions on the Goldstone spectrum and the unfolded level spacing distribution of the Dirac operator at finite density. We compared the Goldstone spectrum 
to predictions from leading order chiral perturbation theory for two sets
of lattice parameters: 
\begin{itemize}
\item A $12^3\times 24$ lattice with $\beta=1.5$ and bare mass
$am=0.025$ using a standard Wilson gauge action. The density of 
$Z_2$ monopoles at $\mu=0$ is $\langle z \rangle \sim 0.88$ in
this case.
\item A $16^3\times 32$ lattice with $\beta=1.7$ and bare mass
$am=0.01$ using a treelevel improved Symanzik gauge action. The density of 
$Z_2$ monopoles at $\mu=0$ is $\langle z \rangle \sim 0.27$ in
this case.
\end{itemize}
Our main result is that the Goldstone spectrum switches from 
that of any-color QCD with adjoint fermions in the bulk phase to that of
twocolor QCD with fundamental quarks on the physical weak-coupling side 
of the bulk crossover, as most notably visible in a change of the pion
branch. We show that this change is reflected in the unfolded level
spacing distribution, which appears to obtain a larger and larger
contribution from the Gaussian orthogonal random-matrix ensemble,
starting with the low-lying eigenmodes, as one moves from strong to
weak compling, while deeply in the bulk phase the distribution
is completely dominated by the Gaussian symplectic ensemble.
We conclude that a continuum limit leading to twocolor QCD 
with the correct chiral symmetry-breaking pattern is possible
with rooted staggered quarks.

We also observed that the standard connected susceptibility subtraction to
obtain a renormalized chiral condensate cannot be used at finite
$\mu$, since the connected susceptibility contains a singular contribution
at the diquark condensation transition and that 
a renormalization using a heavy quark condensate is also rendered unfeasible.
Developing a proper $\mu$-dependent renormalization scheme might be
possible using gradient flow techniques, but is left for future work. 

\begin{acknowledgments}
This work was supported by the Helmholtz International Center for FAIR within the LOEWE initiative of the State of Hesse.
 
\end{acknowledgments}

\newpage
\bibliographystyle{JHEP}
\bibliography{paper-QC2D}

\end{document}